\documentclass[reprint,amssymb,amsmath,aip,cha]{revtex4-2}
\usepackage{graphicx}
\usepackage[caption = false]{subfig}
\usepackage{color}
\usepackage{epsfig}
\usepackage{ifpdf}
\DeclareUnicodeCharacter{2009}{\,}
\DeclareUnicodeCharacter{00D7}{\times}
\usepackage{bm}
\usepackage[colorlinks=true,linkcolor=blue]{hyperref}%
\expandafter\ifx\csname package@font\endcsname\relax\else
 \expandafter\expandafter
 \expandafter\usepackage
 \expandafter\expandafter
 \expandafter{\csname package@font\endcsname}%
\fi
\usepackage{cleveref}
\begin{document}
\title{Magnetized dusty plasma: On issues of its complexity and magnetization of charged dust particles}
\author{Mangilal Choudhary}
\affiliation{Department of Physics and Astrophysics, University of Delhi, Delhi-110007, India}
%
\begin{abstract}
It is possible to excite various linear and non-linear low-frequency modes in dusty plasma which is an admixture of electrons, ions, gas atoms, and negatively charged solid particles. The experimental as well as theoretical study of these low-frequency dynamical modes in dusty plasma is very complex because of the involvement of dynamics of electrons, ions, and neutrals. If the external magnetic field is introduced to dusty plasma then the dynamics of it will be more complex. The complexity of magnetized dusty plasma where plasma species are magnetized is discussed by keeping the experimental observations in magnetized dusty plasma devices in mind. The requirement of theoretical modeling, as well as computation experiments in understanding the dynamics of dusty plasma in the presence of a strong magnetic field, is highlighted in the context of experimental findings. The major challenges to magnetizing charged massive particles in experiments and some expected solutions are discussed in this report.      
\end{abstract}
\maketitle
\section{Introduction}
An ionized gas is assumed to be a plasma state (the fourth state of matter) if it qualifies certain conditions\cite{chenplasmaphysicsbook}. This is a fact that plasma is a complex dynamical state however it becomes a more complex system if solid particles of sized nm to $\mu m$ are added to it. In the background of charged species (electrons and ions), these solid particles undergo various charging processes and acquire either negative or positive charges on their surface. In low-temperature plasma ($T_e >> T_i$), dust grains acquire negative charges up to $10^{-16}$ to $10^{-14}$ C due to the collection of more energetic electrons than the slowly moving ions \cite{Charging,dustcharginggoree}. These well seperated negatively charged solid particles float in the plasma medium and can modify the characteristics response of ambient plasma \cite{diaschokwaves,diawexp1,eedfindusty1} with external or internal electromagnetic perturbation. However, these charged massive particles always settle down on the wall of the experimental device because of the finite gravitational force acting on them. A strong electric field can be created using various discharge configurations \cite{dawmerlino, dcdustyplasma,pdasw,mangilalpop,mangirsiexpsystem,neerajcrystalccp1} for providing confinement to charged particles against gravity in plasma. Once the density of dust grains in the electrostatic confining potential well crosses a threshold density then long-range coulomb interaction among negatively charged dust particles turns the dust grain medium to respond collectively with internal or external perturbation. \\
In last more than 25 years, a wide spectrum of experimental and theoretical studies has been devoted to exploring the collective response of dusty plasma in the form of various dynamical structures/phenomena such as dust acoustic waves\cite{ddw1,dasw,daw3,ddw2,ddwfortov,mangilalpop,pdasw}, rotational and vortex motion\cite{rotationkarasev2,rotationkarasev2,mangilalannulardusty,vortexmicrogravity,mangilalmultiplerot,vikramtsw, mangilargeaspect,modhuvortices,bellanicedustyrotation}, crystallization and phase transition \cite{thomasdustycrystal,linidustycrystal2,dustcrystalhariparasad,melzerdustcrystalmelting}, instability driven modes \cite{sanatkhinstability,vikramkhinstability,rtinstabilityvortices,avinashrtinstability,vikramrtinstability} and voids \cite{void,cavity,ringvoid}. The major objectives of these theoretical and experimental studies of dusty plasma were to understand the naturally occurring dusty plasmas, control the impurities in semiconductor industries, explore the role of impurities on characteristics of plasma, establish it as a model system to understand the complex physical systems, etc. After a wide range of dusty plasma studies, still there are some open questions: Could we understand all astrophysical systems (plasma with dust impurities) with existing experimental results or available theoretical models? Do we need to explore the dusty plasma in the presence of a magnetic field? Would be there a finite role of magnetic field on the cloud of charged dust grains? Getting the answers to such open questions is only possible if dedicated experimental and theoretical studies of magnetized dusty plasma will be conducted.\\
In recent years, magnetized dusty plasma has been a popular research topic among the dusty plasma community worldwide. The dynamics of dust grain medium or well-separated dust grains strongly depends on the characteristics of background plasma that can be modified in the presence of the external magnetic field. It is expected that the complexity of dusty plasma increases in the presence of external magnetic field. The magnetization of electrons is possible at a low magnetic field (B $< 0.05 T$) as compared to massive charged dust particles (B $>$ 2 T). The study of dusty plasma in the presence of an external magnetic field (magnetized dusty plasma) is categorized as weakly and strongly magnetized dusty plasma. A wide range of theoretical as well experimental work on weakly magnetized dusty plasma where only electrons are magnetized is conducted by various research groups\cite{weaklymagnetized1,weaklymagnetized2,weaklymagnetized3}. It is a fact that strong magnetic field is required for the study of strongly magnetized dusty plasma. There are few devices based on permanent magnet \cite{stronglymagnetizeddc1} and superconducting electromagnet \cite{thomasmpedx,melzermagnetizeddusty,mangilaljpp} for introducing the strong magnetic field (few Tesla) to dusty plasma. The researchers are continuously putting effort to explore the effect of magnetic field on dusty plasma phenomena such as excitation of waves, charging mechanism, rotational and vortex motion, crystallization and melting, etc. However, there are many challenges in understanding the experimentally observed dynamics of dusty plasma because of its complex nature.\\ 
A detailed discussion on a few issues regarding understanding the experimental results and producing strongly magnetized dusty plasma is given in subsequent sections. The complexity of magnetized dusty plasma is discussed in Sec.~\ref{sec:secI}. The major challenges in achieving magnetization conditions of negatively charged dust grains in rf discharges at strong B-field are discussed in Sec.~\ref{sec:secII}. The proposed suggestions to get magnetized dusty plasma (dust particles will be magnetized) are given in Sec.~\ref{sec:secIII}. The concluding remark is given in Sec.~\ref{sec:secIV}.
\section{Complexity of magnetized dusty plasma}  \label{sec:secI}
Dusty plasma is considered a complex dynamical system. It becomes a more complex system once the external magnetic field is applied. The complexity of weakly or strongly magnetized dusty plasma in terms of various experimentally observed physical phenomena is discussed in respective subsections.    
\subsubsection{Dust charging}
In low-temperature weakly ionized plasma, charging of solid dust particles mainly depends on the collection currents of plasma species \cite{chargingbarken,dustcharginggoree,shukladusty3}. The magnitude of electron current ($I_e$) and ion current ($I_i$) to dust surface strongly depends on their density and average energy at the fixed neutral background. The charging currents are expected to change once the magnetic field is introduced to dusty plasma. As the magnetic field is applied, the gyro-radius of electrons ($r_{ge} = m_e v_{te}/e B$) and of ions ($r_{gi} = m_i v_{ti}/e B$) decreases with increasing the strength of external B-field. The electrons are magnetized at a lower magnetic field (B $<$ 0.05 T) but a higher B-field (B $>$ 0.1 T) is required to magnetize ions in typical dusty plasma experiments \cite{mangilaljpp}. In magnetized plasma, the currents $I_e$ and $I_i$ to the dust surface are expected to change due to the following reasons: (a) magnetic field reduces the loss of energetic electrons from the plasma region to the wall of the experimental chamber. The confinement of energetic electrons may change the average energy or electron temperature of bulk plasma electrons \cite{mangilaljpp}. This change in electron temperature may alter the charging currents. (b) Magnetic field (weak) may increase the confinement of electrons \cite{mangilaljpp} which could be a cause of the change in the charging mechanism. (c) Magnetic field modifies the collisional frequency of electrons/ions with background neutrals. In this way, the ionization rate may get modified in the presence of the magnetic field. Modification in the plasma formation rate or ionization process will affect the charging currents. (d) In magnetized plasma, net electron current ($I_e$) is assumed to be the sum of two possible currents along B-field ($I_{e\parallel}$) and transverse to B-field ($I_{e\perp}$). The $I_{e\perp}$ strongly depends on the strength of the external B-field which definitely will affect the charging currents. (e) Once ions are magnetized then net ion current to the dust surface will have two components along B-field ($I_{i\parallel}$) and transverse to B-field ($I_{i\perp}$. The change in ion current may increase or decrease the dust charge at strong B-filed\cite{chenplasmaphysicsbook,tsytovichdustcharge1}. (f) It is also expected to observe a finite effect of collisions (background neutral density) on the charging currents in the presence of the magnetic field due to the $I_{e\perp}$ and $I_{i\perp}$ dependence on collisional relaxation time \cite{chenplasmaphysicsbook,mangilaljpp}. 
\\\\
It should be noted that the estimation of accurate charge on dust grains in weakly/strongly magnetized plasma is necessary to understand the dynamics/characteristics of dusty plasma. In the last few years, theoretical as well as experimental works have been carried out to estimate the charge on dust grains in magnetized plasma. Melzer \textit{et al.} \cite{melzerdustchargeb} adopted normal mode analysis of the dust cluster in the magnetized rf plasma for extracting charge on nm to micron-sized dust grains. At the same time, Choudhary \textit{et al.} \cite{mangilaljpp} estimated charge on magnetic and non-magnetic spherical probes (large dust grains) in magnetized rf discharge. The theoretical study and computational experiments have been performed to understand the charging mechanism in magnetized plasma background \cite{tsytovichdustcharge1, langefloatinginmagnetized,dustcurrent,tomitadustchargingwithmagneticfield,kodanovadustchargeb,picsimulationmagnetizedcharge} 
However, there are inconsistencies in the numerically estimated and experimentally observed values of the dust charge in weakly magnetized dusty plasma. The available theoretical models to understand the charging mechanism of non-magnetic (magnetic) solid spherical particles in the magnetized plasma (weakly and strongly) are incomplete. Therefore, an adequate theoretical model and novel experimental techniques are required to estimate the accurate charge on non-magnetic and magnetic solid particles in the background magnetized plasma at different strengths of magnetic fields. 
\subsubsection{Dust-acoustic waves}
\begin{figure} 
\centering
\includegraphics[scale= 0.4500000]{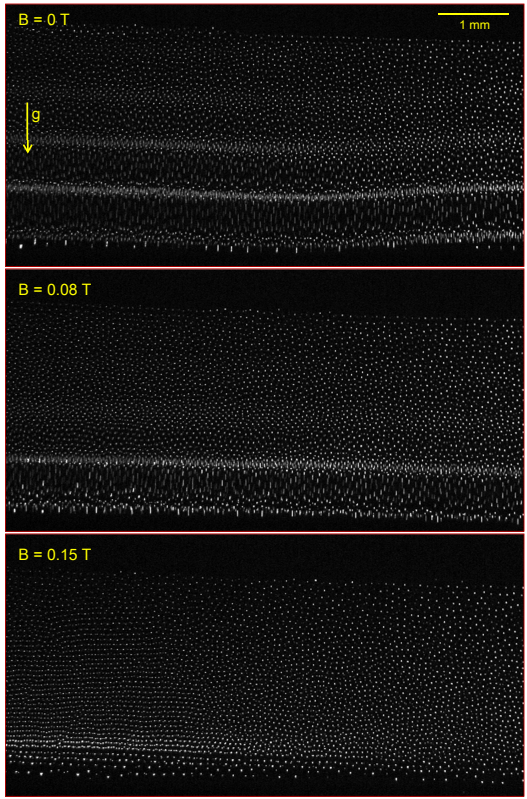}
\caption{\label{fig:fig1} Propagation of dust acoustic waves at different strengths of external magnetic fields. Details of experimental configuration and parameters are available in ref.\cite{mangilalmagneticdaw}}
\end{figure}

The excitation of low-frequency modes in dusty plasma is a result of its collective response in the presence of internal/external electromagnetic perturbation. The study of dust acoustic waves (low-frequency modes) has a more than 25 years old history \cite{dawmerlino} and is still a vibrant research topic for researchers. The propagation characteristics of dust acoustic modes in low-temperature direct current (DC) and radio-frequency (rf) plasma \cite{mangilalpop,pdasw,daw2,rfdischarge,mangirsiexpsystem,icpddw,tsw,dlw2} were explained based on the available theoretical models \cite{raodaw1,dasw,shukladusty3,dlw1}. Recently, Choudhary \textit{et al.} \cite{mangilalmagneticdaw} and Melzer \textit{et al.} \cite{melzerdaw} conducted experimental studies on self-excited dust acoustic waves (DAWs) in the presence of external magnetic field and observed the damping as well modification in propagation parameters at the higher magnetic field. The role of external magnetic field on propagating DAW in capacitative coupled discharge \cite{mangilalmagneticdaw} is shown in Fig.~\ref{fig:fig1}. The propagating waves are getting damped with increasing the strength of external B-field. Thus we do not see any wave modes in dusty plasma medium at strong B-field (see Fig.~\ref{fig:fig1}).\\  
The excitation and propagation characteristics of dust acoustic waves strongly depend on the background electric field, neutrals density, dust charge, plasma density, etc. Since the magnetic field modifies plasma parameters, the propagation characteristics of DAWs get modified with the application of external B-field. Salimullah \textit{et al.}\cite{dawmagnetized1} performed a theoretical study of DAWs in the magnetized plasma. However, this proposed theoretical model could not explain the modification and damping of waves which were observed experimentally in magnetized dusty plasma devices\cite{mangilalmagneticdaw,melzerdaw}. The interpretation of experimentally observed results on modification and damping of DAWs in the magnetized plasma could only be possible if an adequate theoretical model will be developed.
\subsubsection{Rotational motion and vortex formation} 
\begin{figure*} 
 \centering
\subfloat{{\includegraphics[scale=0.550050]{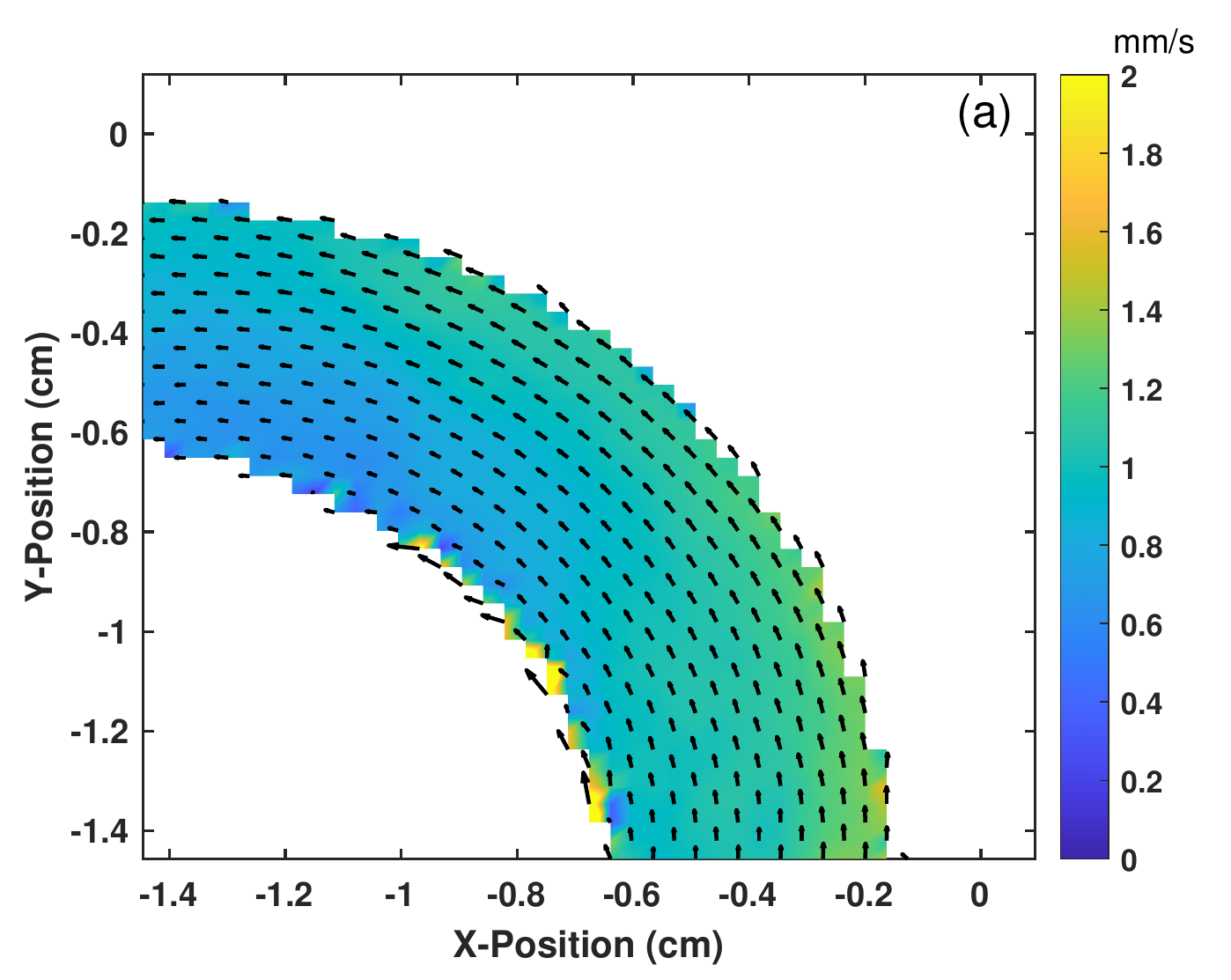}}}%
\hspace*{0.1in}
 \subfloat{{\includegraphics[scale=0.55050]{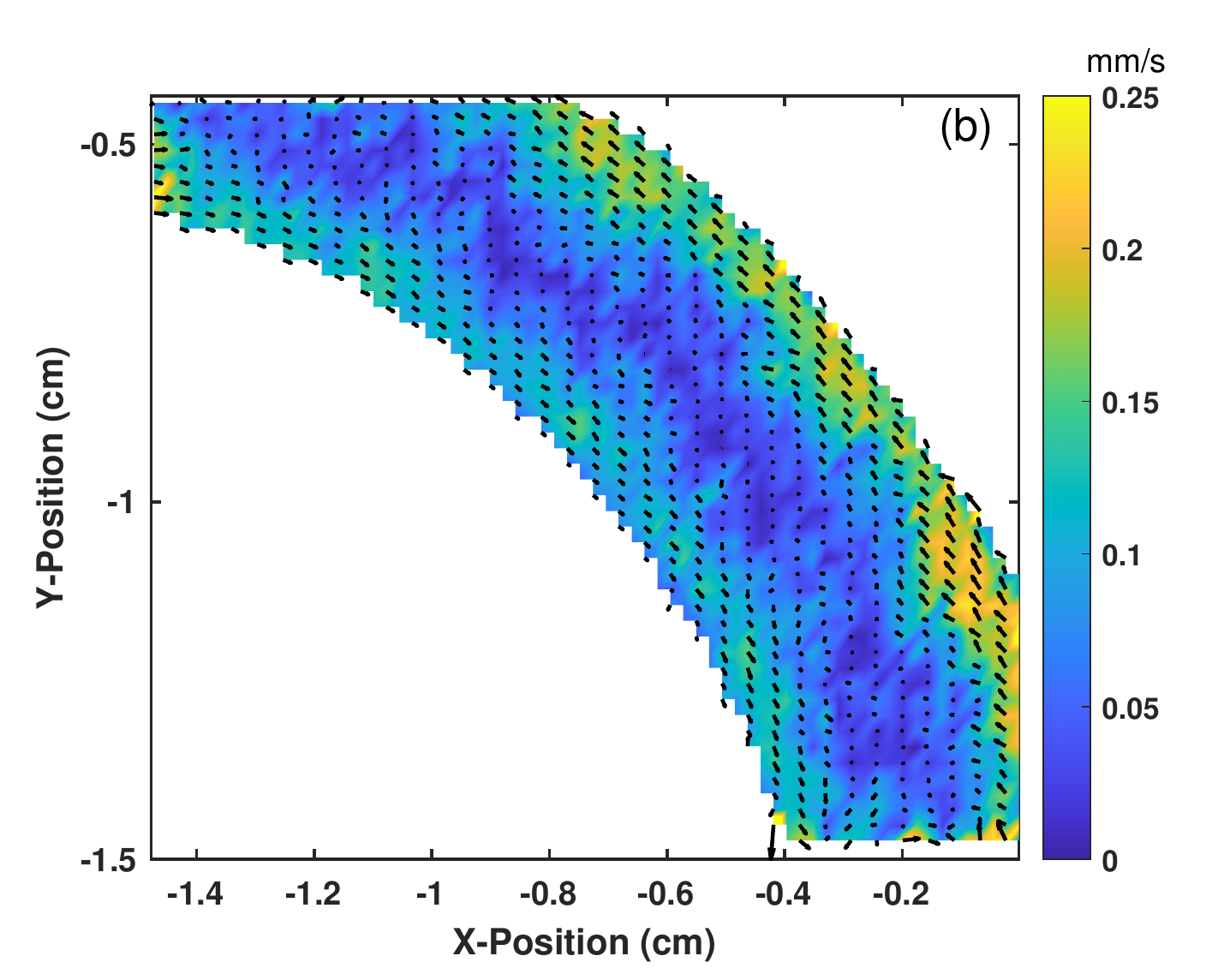}}}
 \quad
\caption{\label{fig:fig2}(a)PIV image (average of 50 images) of annular dusty plasma flow (uni-directional) between two conducting (aluminum) rings (inner and outer diameter of the first ring are 5 mm and 12 mm respectively, same for the second ring are 30 mm and 50 respectively).  The diameter of dust particles is 6.28 $\mu$ m and argon pressure during experiments was 30 Pa. The peak-to-peak rf voltage was 55 V and the applied B-field was 0.2 T. (b) PIV image of annular dusty plasma flow (opposite-rotation) between two non-conducting (Teflon) rings (inner and outer diameter of the first ring are 20 mm and 30 mm, the same for the second ring are 40 mm and 50 mm respectively). The diameter of dust particles is 6.28 $\mu$ m and argon pressure during experiments was 30 Pa. The peak-to-peak voltage was 55 V and the applied B-field was 0.4 T. More details about the experimental configuration to create annular dusty plasma are given in ref.~\cite{mangilalannulardusty}} 
\end{figure*}

In the absence of magnetic field, dusty plasma shows rotational motion as well as vortex flow \cite{agarwalrotation,manjeetrotation,bellanicedustyrotation,thermalcreeprotation}. The collective dynamics of 2-dimensional (2D) dust grains medium in the presence of magnetic field where medium exhibits rotational motion is studied by many researchers \cite{rotationknopka1,rotationkarasev2,clusterrotationunmagnetizedplasma,inductivelycoupledrotation,rotationinionflow}. In presence of magnetic field, 3-dimensional dusty plasma exhibits vortex motion and possibility to form counter-rotating vortices in a vertical plane \cite{mangimagneticrotation,circulation}. There are available theoretical models to explain the onset of rotational motion \cite{kawmodelrotation} and vortex flow \cite{selfexcitedmotioninhomogeneus,bellanicedustyrotation,laishramshearflowexplaination,modelingdustvortices} in unmagnetized and magnetized dusty plasma. Choudhary \textit{et al.} \cite{mangilalannulardusty} carried out the study of 2D annular dusty plasma in the presence of a strong magnetic field. The dust grains rotate anti-clockwise (uni-directional) in the presence of B-field as shown in Fig.~\ref{fig:fig2}(a). It is expected (as per experimental configuration) to observe the opposite flow of dust grains in the annular region in the presence of B-field.   
However, the opposite flow of dust grains in the annular region in the presence of B-field is only possible with specific discharge conditions. A typical PIV image of an annular dusty plasma at a magnetic field (B = 0.4 T) where dust grains rotate in opposite directions is shown in Fig.~\ref{fig:fig2}(b). Strong theoretical support is required to obtain the specific discharge conditions for generating the unidirectional or opposite flow of dust particles in annular magnetized dusty plasma.\\
%
\begin{figure*} 
\centering
\includegraphics[scale= 0.8500000]{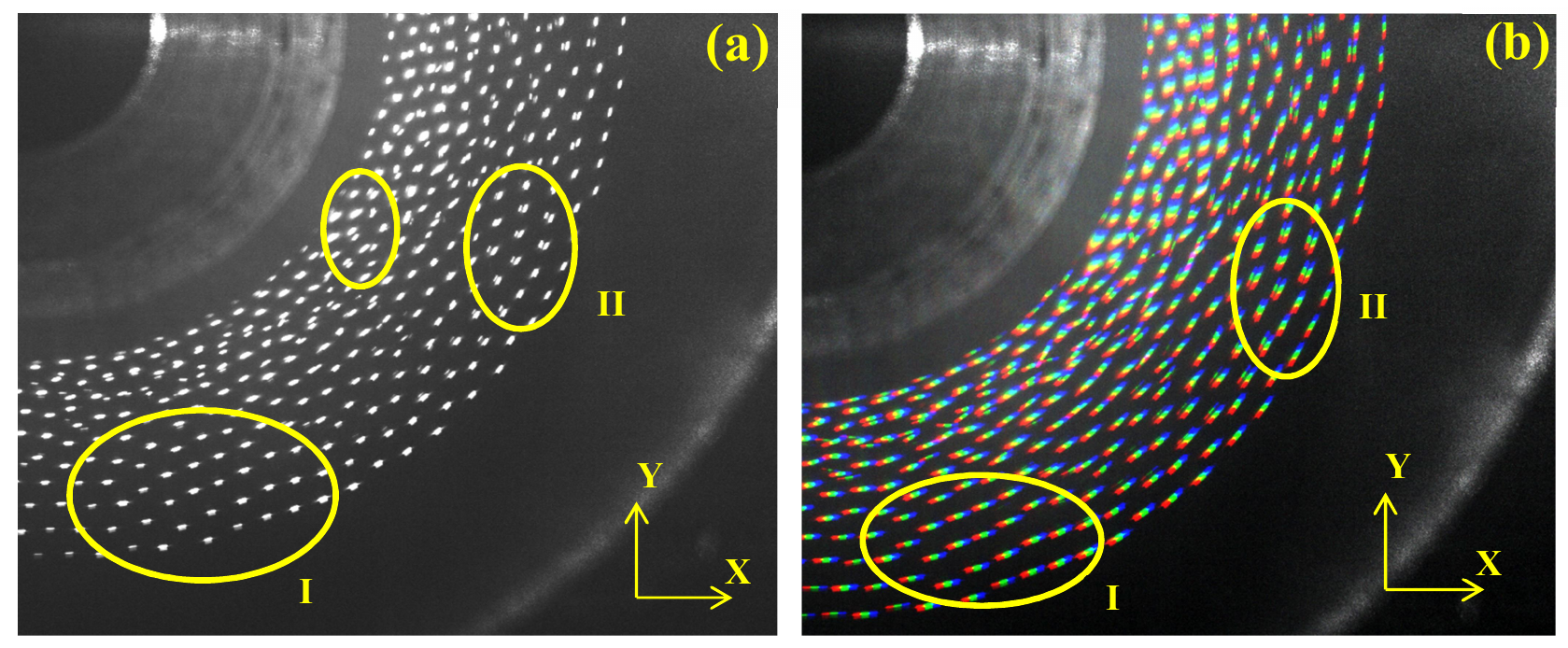}
\caption{\label{fig:fig3} (a) A dusty plasma image in a horizontal plane that shows the pairing of dust grains in an annular region at a strong magnetic field (B = 0.6 T). There is no pairing of dust particles in the encircled region I and a clear pairing of two dust particles in encircled region II. (b) A composite image of three still images taken at different times (dt = 80 ms). The elongated dust tracks in the encircled region I represent the motion of single dust grains. Various elongated pairs in encircled region II clearly indicate the motion of pairs up to a certain distance with time. The experiments were performed in argon gas at 30 Pa. The annular dust medium was created in between two conducting (aluminum) rings (the inner and outer diameters of the first ring are 5 mm and 12 mm respectively, same for the second ring are 30 mm and 50 respectively).  The diameter of dust particles is 6.28 $\mu$ m and peak-to-peak rf voltage was 55 V}
\end{figure*} 
At a strong magnetic field, dust grains in an annular region try to be in a pair and flow together up to a certain distance. A still image of annular dusty plasma is shown in Fig.~\ref{fig:fig3}(a) in which encircled regions (I and II) highlight the isolated and pair of dust particles, respectively. Three images at different times are superimposed to see the motion of isolated particles and dust pairs in the B-field. Encircled regions (I and II) in Fig.~\ref{fig:fig3}(b) show the elongated tracks of isolated dust particles and dust pairs during the rotational motion. It is required to develop a theoretical model for understanding such complex phenomena of annular dusty plasma in the presence of strong magnetic field.\\ 
The recent experimental study of 3-dimensional dusty plasma in a strong magnetic field \cite{mangimagneticrotation} (B $>$ 0.1 T) demonstrates its complexity due to the formation of rotating dust torus (vortices in the vertical plane) and rotation of dust particles in the horizontal plane. A typical image of dusty plasma in the vertical and horizontal plane is shown in Fig.~\ref{fig:fig4}. The dust grains are confined in a bowl-shaped potential well and the dynamics of grains can be tracked by analyzing images either in a vertical plane (see Fig.~\ref{fig:fig4}(a)) or horizontal plane (see Fig.~\ref{fig:fig4}(b)). It is observed that the dynamics of dust grains become more complex in the presence of B-field if the volume of the medium is increased. A large volume (3D) dusty plasma can be created by using smaller-sized dust particles in the experiment. The dynamics of 3D dusty plasma at different B-fields are demonstrated by PIV images of the medium in a horizontal plane at different positions in vertical directions (see Fig:~\ref{fig:fig5}). The topmost dust layers show only rotational motion in the azimuthal direction. In between the topmost and bottom layers, dusty plasma exhibits mixed motion (rotation in the horizontal plane and vortex flow in the vertical plane) in the presence of a strong B-field. The direction of vortex flow (torus in 3D dusty plasma) changes (anti-clockwise to clockwise) as we move towards the bottom layer, as displayed in  Fig:~\ref{fig:fig5}. The results (piv images) indicate the formation of multiple counter-rotating tori in the large-volume magnetized dusty plasma. The formation and direction of vortex flow along with the rotational motion can be explained with the help of an appropriate futuristic theoretical model.   
\begin{figure*} 
\centering
\includegraphics[scale= 0.8500000]{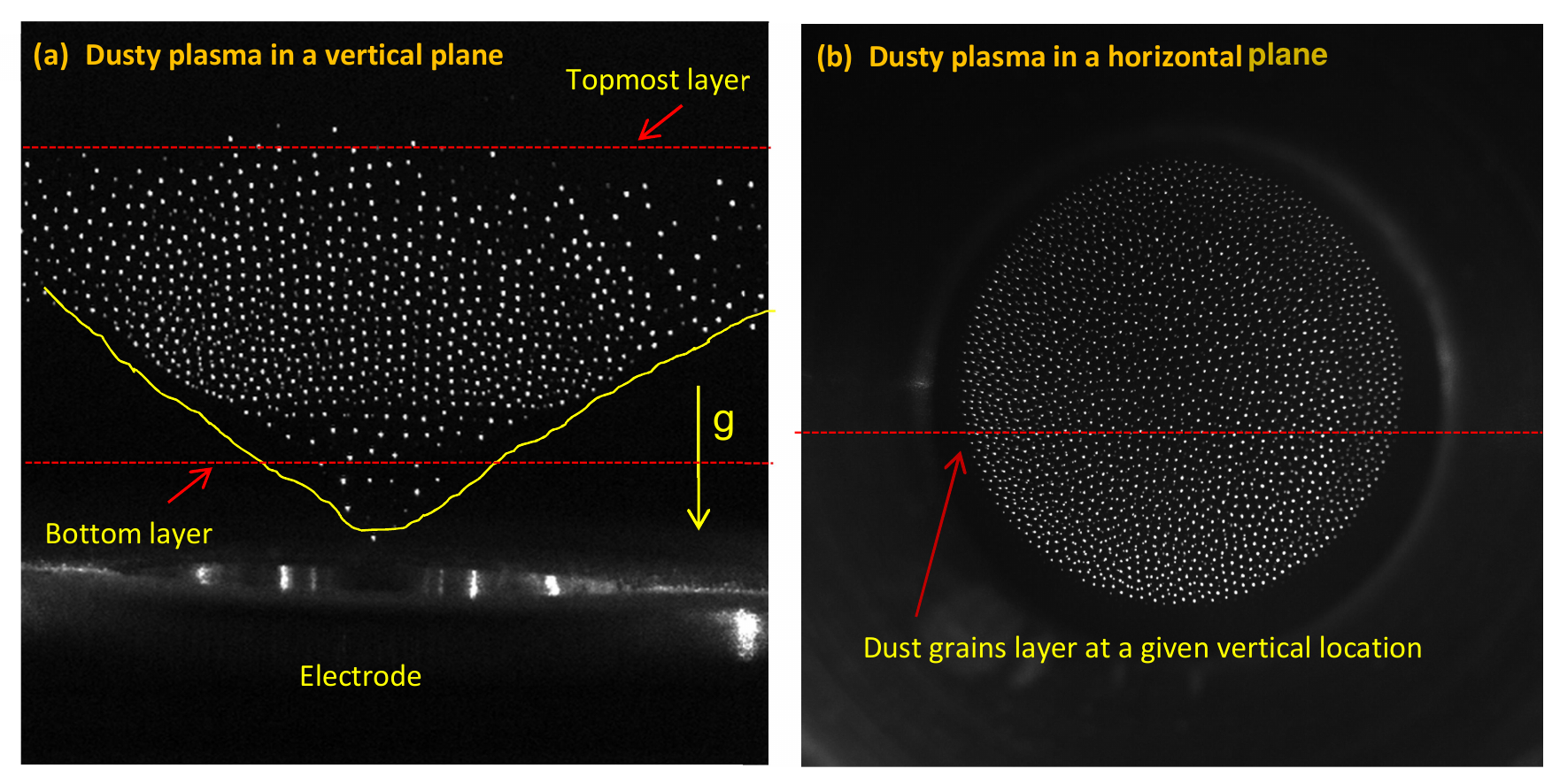}
\caption{\label{fig:fig4} Typical dusty plasma in a vertical (Y--Z) plane (b) in horizontal (X--Y) plane. The dust grains are confined in a potential well created by placing an addition of either conducting or non-conducting ring of an appropriate size and thickness. The 3D dusty plasma looks like a bowl shape in the vertical plane, as denoted by the boundary. The width of dusty plasma in the vertical direction depends on the depth of the potential well and the size of dust particles}
\end{figure*} 
\begin{figure*} 
 \centering
\subfloat{{\includegraphics[scale=0.550050]{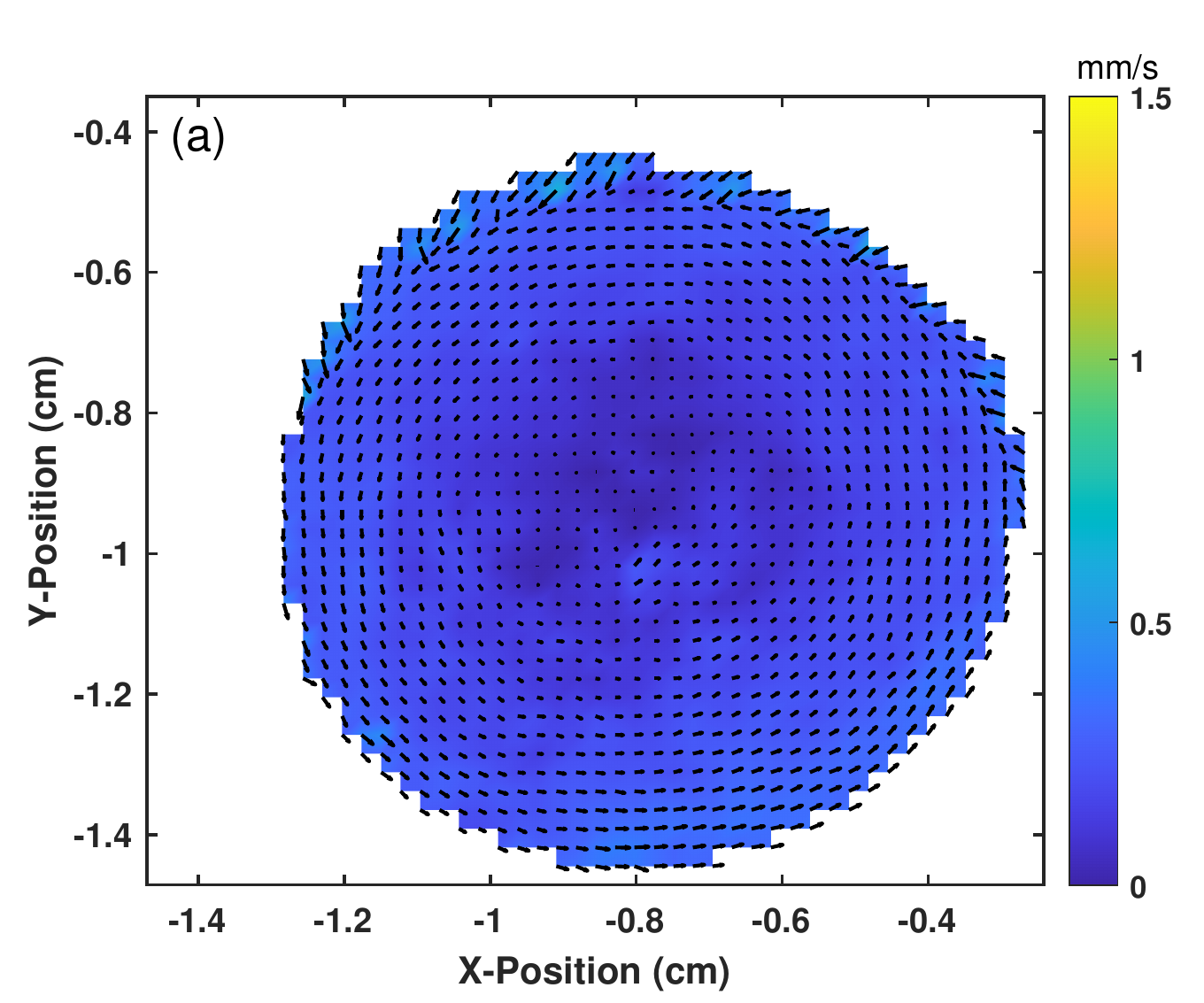}}}%
\hspace*{0.1in}
 \subfloat{{\includegraphics[scale=0.55050]{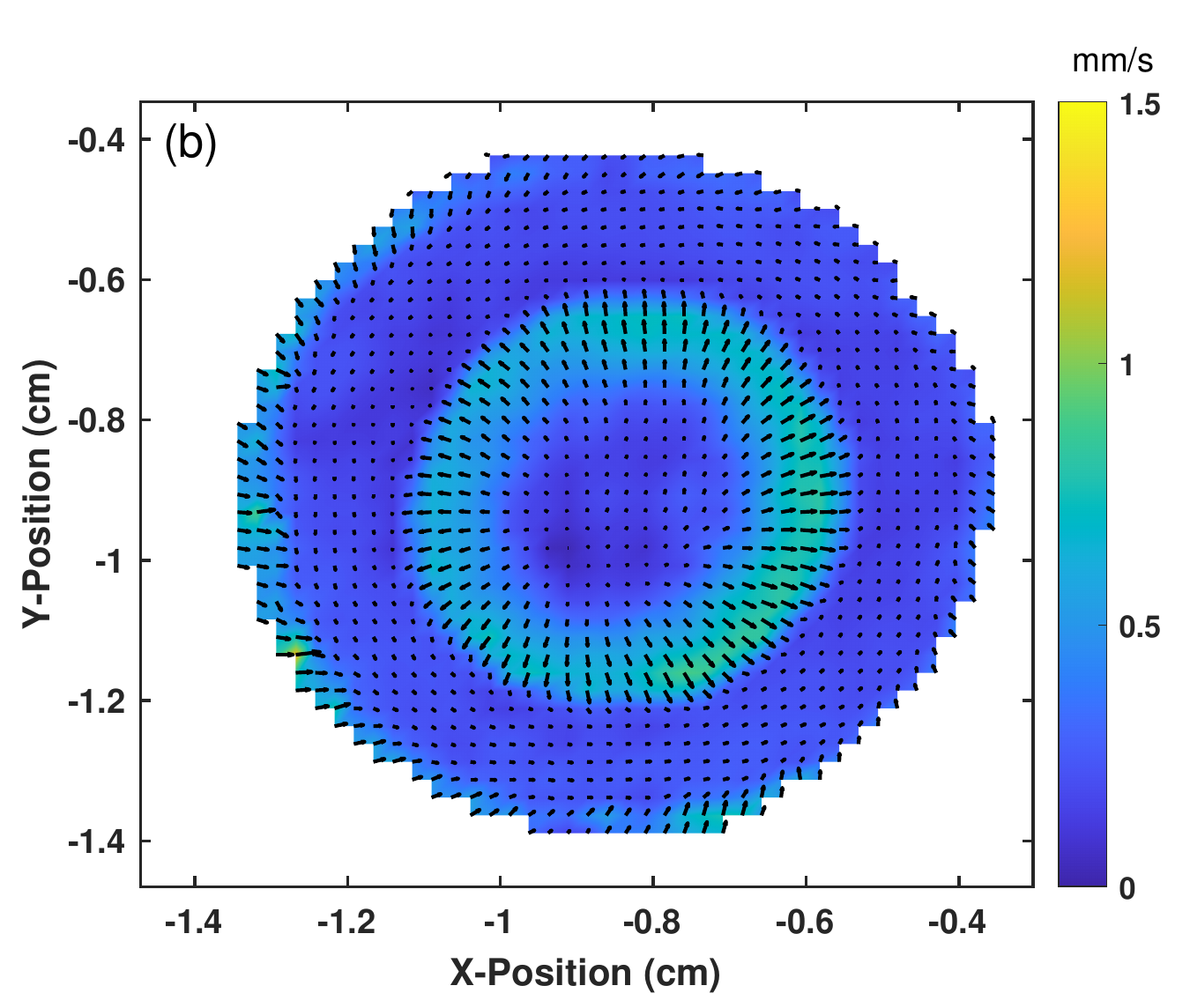}}}
 \quad
 \hspace*{0.1in}
 \subfloat{{\includegraphics[scale=0.55050]{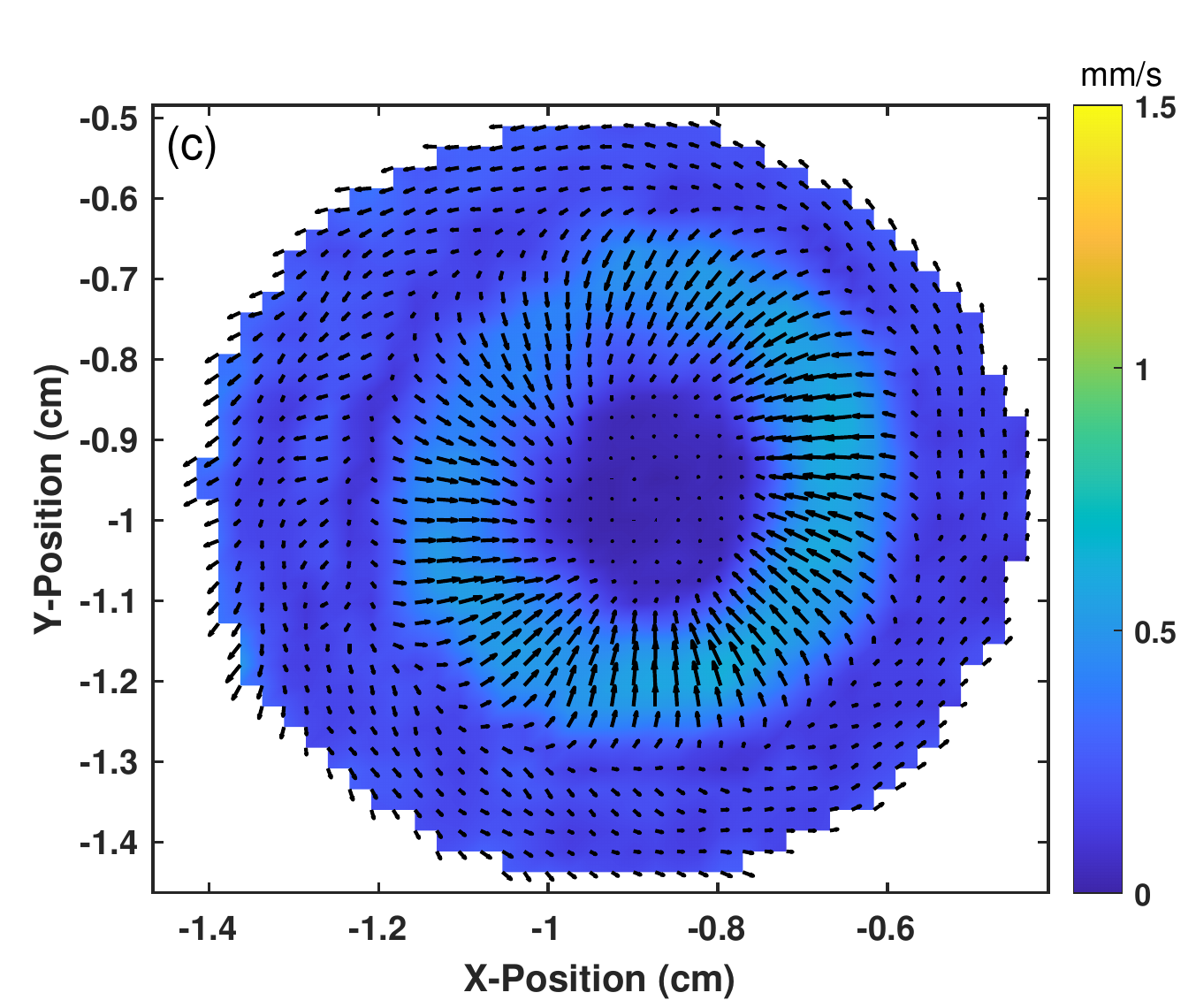}}}
 \hspace*{0.1in}
 \subfloat{{\includegraphics[scale=0.55050]{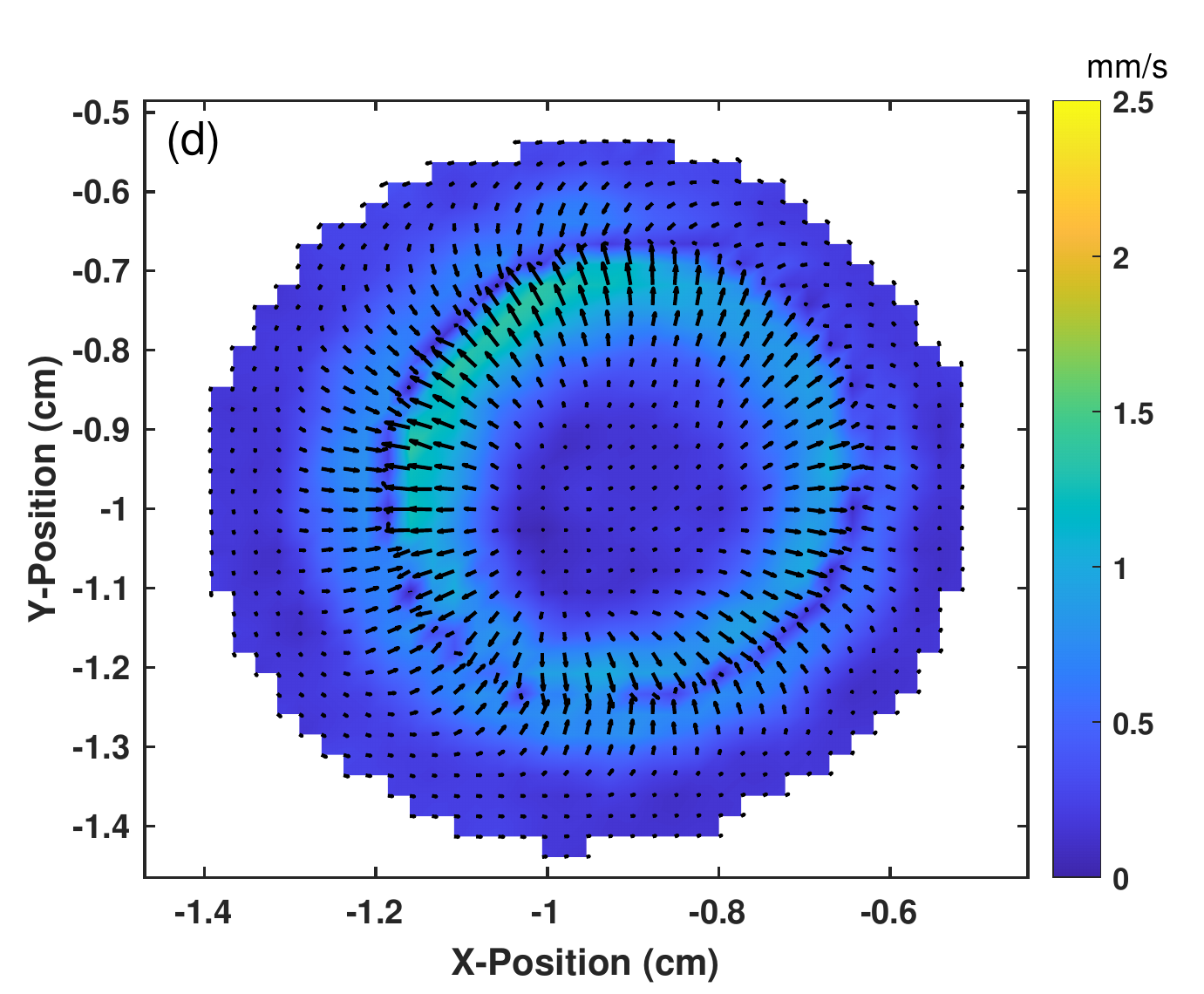}}}
 \quad
 \hspace*{0.1in}
 \subfloat{{\includegraphics[scale=0.55050]{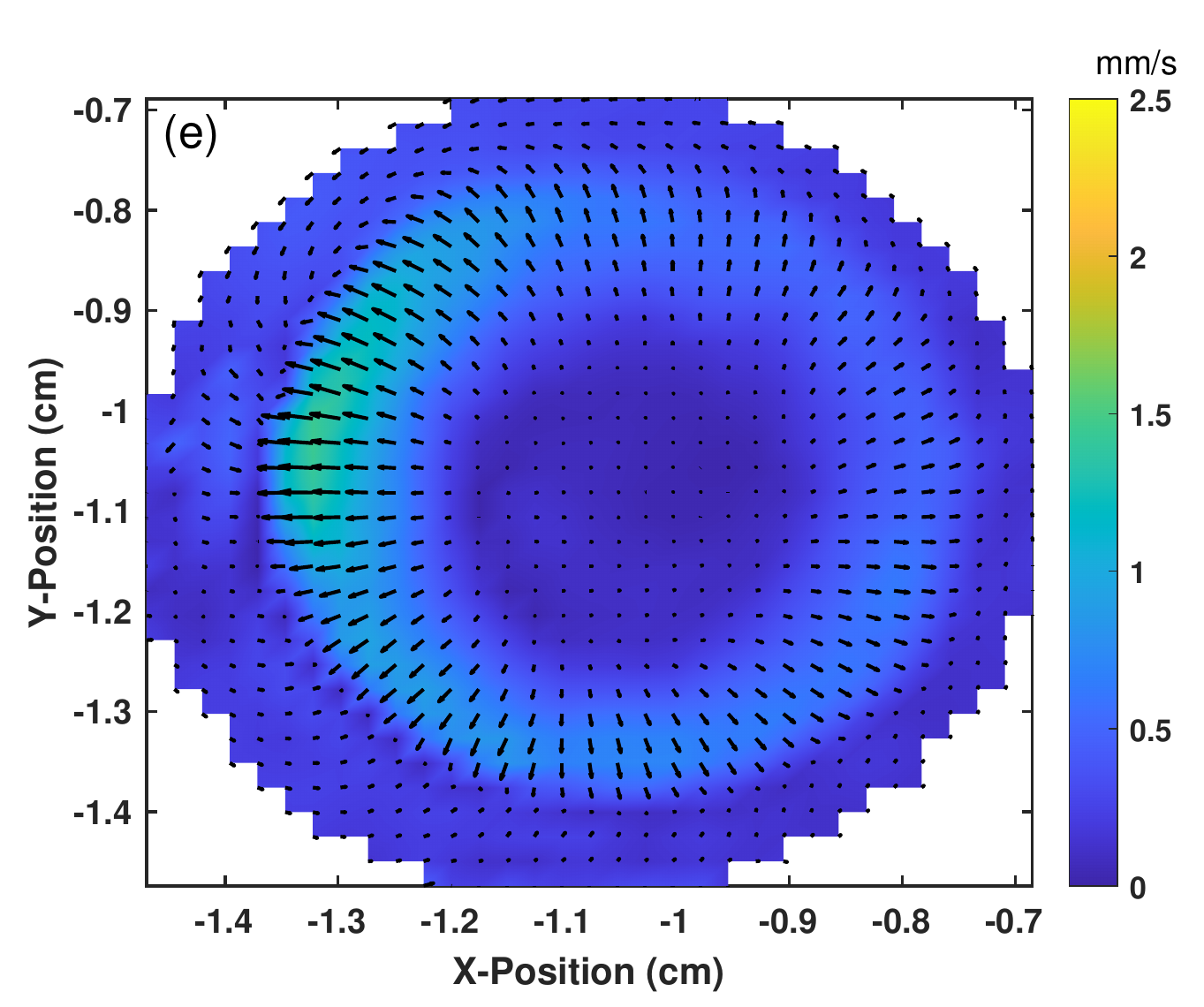}}}
 \quad
 \hspace*{0.1in}
 \subfloat{{\includegraphics[scale=0.55050]{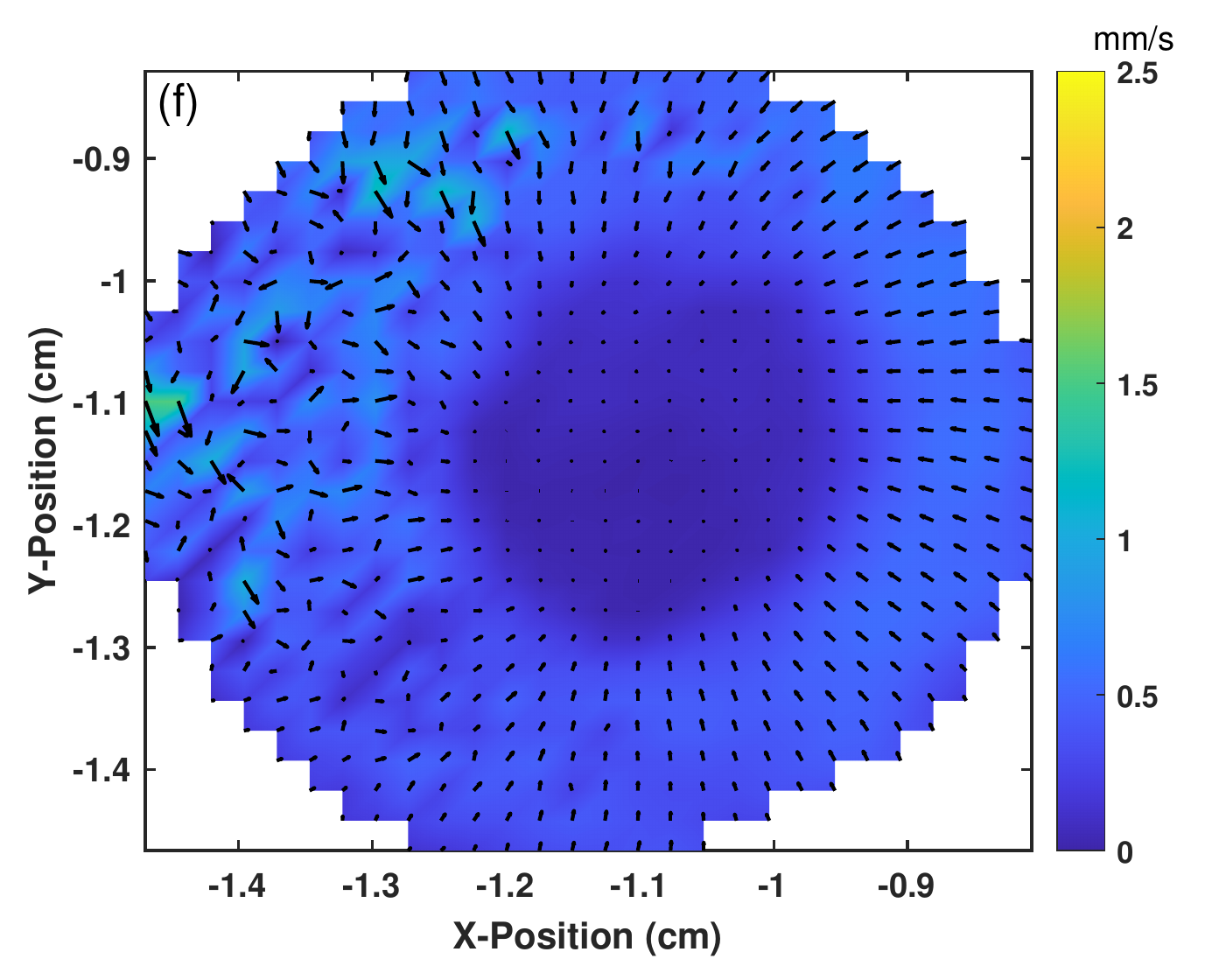}}}
 \quad
\caption{\label{fig:fig5}(a)PIV images of dusty plasma (obtained after averaging 50 images) flow in a horizontal plane at different vertical locations of confined dust grains in potential well (a) Topmost dust grains (b) 2 mm below the topmost dust grains plane (c) 4 mm lower the topmost dust grains plane (d) 6 mm lower the topmost dust grains plane (e) 8 mm lower the topmost dust grains plane (f) 12 mm lower the topmost dust grains bed/layer. The diameter of dust particles is 2 $\mu$ m and argon pressure was set at 35 Pa. The peak-to-peak rf voltage was 52 V and the applied magnetic field was 0.15 T} 
\end{figure*}
\begin{figure} 
\centering
\includegraphics[scale= 0.4500000]{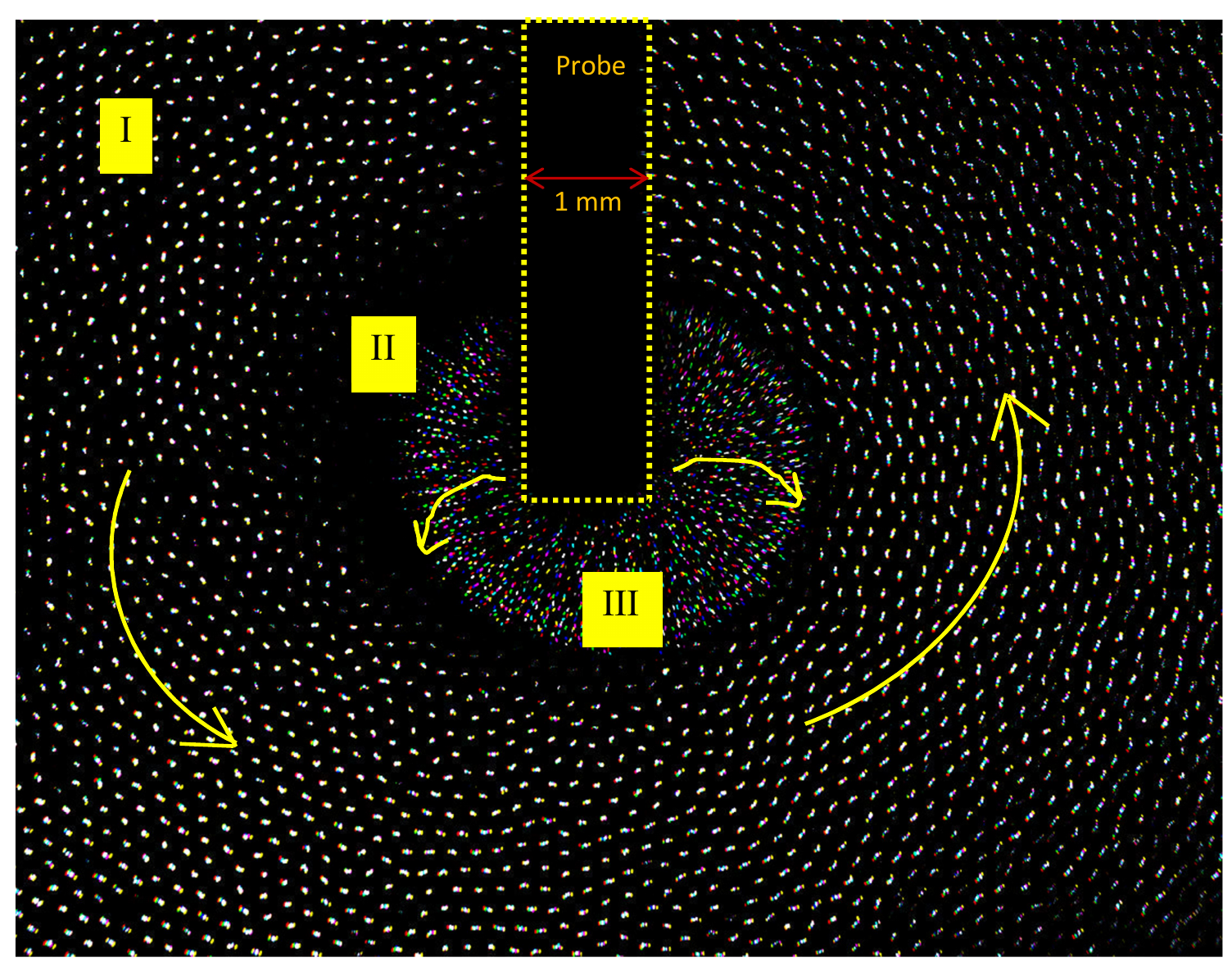}
\caption{\label{fig:fig6} A composite image of five still images at different time intervals (dt = 120 ms) in the presence of a positively biased (+70 V) cylindrical probe (in a vertical direction) in a 3-dimensional magnetized dusty plasma. Three different regions are marked in the image. Region I represents the dust grain medium (2D image) far away from the biased probe, region II represents the dust-free region (void) and region III indicates the outward motion of dust grains which is a signature of the 3D motion of dust particles. The experiments were performed in argon plasma at a pressure of 40 Pa. The peak-to-peak voltage on the electrode is 55 V and the external applied B-field is 0.7 T.}
\end{figure} 
\subsubsection{Void formation}
A dust-free region (void) in dusty plasma is formed either by internal instabilities \cite{microgravityvoid} or external electrostatic perturbation \cite{void}. If a floating or negatively biased cylindrical probe (rod) is inserted into the dusty plasma medium, a dust-free region is formed around it \cite{void,cavity,ringvoid}. The studies on void formation in dusty plasma were carried out either in DC discharge or RF discharge. In unmagnetized equilibrium dusty plasma, dust particles exhibit random motion about their equilibrium position. The formation of void around a negatively biased or floating probe (rod) is the result of competition between inward ion drag force and outward electric field on dust grains near the probe \cite{voidstheorygoree,void}
The experimental results of the dust-free region around negatively biased or floating objects (probe/rod) in unmagnetized dusty plasma can be explained by the available theoretical model \cite{voidstheorygoree,void}. In the presence of the magnetic field, dust particles rotate in a horizontal plane (in the case of 2D dusty plasma) due to the azimuthal flow of ions \cite{rotationinionflow,rotationknopka1,mangilalannulardusty}. The flowing ions set dust particles into rotational motion in a plane perpendicular to the magnetic field. A dust-free region is expected to form around a negatively charged/floating probe (rod) if it is inserted into the rotating dusty plasma. However, the available theoretical models for unmagnetized dusty plasma \cite{void,voidstheorygoree} may need some corrections due to the following reasons: (a) The ion flow in the radial electric field of charged probe (rod) will be modified in the presence of B-field. Only the radial velocity component of ions will be responsible to compensate the electric force acting on dust particles. (b) The electric and ion drag forces acting on dust grains may get changed due to the variation of dust charge \cite{voiddependsonionization} in the presence of B-field. Hence, the characteristics of void may get changed in the magnetized dusty plasma. Thus the theory of void formation in weakly or strongly magnetized dusty plasma would be very complex.\\
The interaction of a positively biased probe (cylindrical/spherical) with 3-dimensional dusty plasma in the presence of B-field will be very complex. A typical image of dust grain medium in the horizontal plane around a cylindrical probe (1 mm diameter) at B = 0.7 T is shown in Fig.~\ref{fig:fig6}. Around the positively biased probe (+70 V), we observed a dust-free region in rotating dusty plasma (plane perpendicular to B-field) and vortex motion of dust particles near the probe in the vertical plane (along the axis of the probe). The complexity in understanding such experimental results of magnetized dusty plasma can be resolved by developing promising theoretical models and computational experiments.  
\subsection*{Crystallization and melting of dusty plasma}
\begin{figure} 
\centering
\includegraphics[scale= 0.4500000]{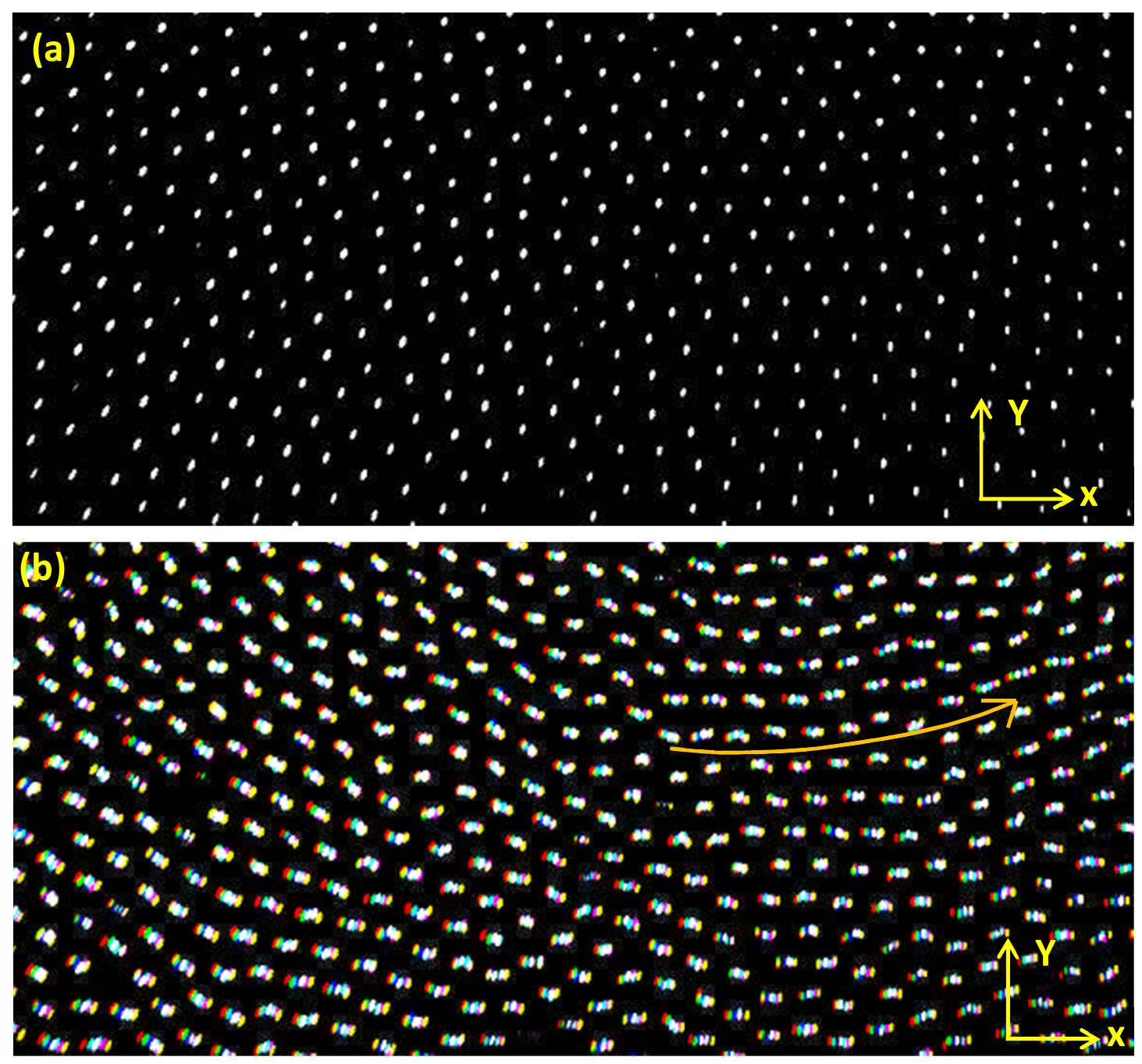}
\caption{\label{fig:fig7} (a) Image shows the crystalline (solid) state of dusty plasma at B = 0 T (b) A composite image of seven still images at different time intervals (dt = 160 ms) in presence of B-field (B = 0.5 T). The dust grains were confined by a smaller aluminum ring electrode (10 mm inner and 20 mm outer diameter, 2 mm width) on the lower powered rf electrode. The size of dust grains was 6.28 $\mu$ m and argon pressure was 30 Pa for this experimental result}
\end{figure}

The study of dusty plasma crystals in various discharge configurations has been a hot research topic in the dusty plasma because of its wide scope in understanding mediums at kinetic level \cite{thomasdustycrystal,crystaldonko2009,linidustycrystal2,dustcrystalhariparasad,crystalarumugamdc2021,crystalneeraj2023}. Apart from neutral pressure and input power, the external magnetic field can also alter the characteristics of dust-plasma crystal \cite{crystalmagneticfieldsurbhi2017}. The dust grains in a crystalline state perform random motion about their equilibrium position in the absence of an external B-field but they have rotational motion if the magnetic field is introduced. A typical still image (crystalline state) and superimposed image of five consecutive images at different times of dusty plasma crystal in the magnetic field of strength 0.5 T is depicted in Fig.~\ref{fig:fig7}. The dust grain medium can be in a  crystalline state (solid) or liquid state or gaseous state while it has rotational motion. We know that coupling constant (strength) and screening length \cite{thomasdustycrystal,crystaldonko2009,dustcrystalhariparasad} decide the state of dusty plasma. Both parameters of dust grain medium are strongly dependent on background plasma parameters, therefore, the external magnetic field definitely can modify the characteristics of dusty plasma crystal. However, there is a lack of theoretical model/computational simulation to understand the dusty plasma states in the presence strong magnetic field.    
\section{Challenges in magnetizing the charged dust particles in strong B-field} \label{sec:secII}
In recent years, dedicated experimental devices are built to study the strongly magnetized dusty plasma where massive charged particles (nm to $\mu$m) can be magnetized \cite{thomasmpedx,mangilaljpp,melzermagnetizeddusty,stronglymagnetizeddc1}. However, there are many challenges to magnetizing the solid-charged dust particles in the ambient plasma background after application of a strong B-field. Thomas \textit{et al.} \cite{thomasmagnetizationdust} suggested the strength of magnetic field (few Tesla) to magnetize the charged dust particles (nm to $\mu$m) in the plasma background. The magnetic field, mass of dust particles, the number density of neutrals (gas pressure), velocity of dust particles, and charge of dust particles are major factors that directly affect the condition of dust magnetization \cite{thomasmagnetizationdust,melzermagnetizationpaper}. The magnetization of dust particles (100 nm to 10 $\mu$m) of low mass density (1 to 1.5 gm/$cm^3$) and low thermal velocity (10 mm to 20 mm/s) is only possible at the high magnetic field (1$<$ B $<$ 5 T) in moderate collisional plasma. It should be noted that theoretical estimation of dust magnetization \cite{thomasmagnetizationdust} in this range of size and velocity is only possible if plasma (dust) parameters remain constant with increasing external B-field. In experimental magnetized dusty plasma, the charging mechanism as well as forces acting on dust grains are strongly dependent on B-field \cite{mangilaljpp,melzerdustchargeb}. There are a few major challenges in achieving the magnetization conditions \cite{thomasmagnetizationdust} for 100 nm to 10 $\mu$m particles in ground-based dusty plasma experiments.
\begin{itemize}
\item The charging mechanism of dust particles strongly depends on the external magnetic field. The charge accumulated on dust grains is expected to change which needs to be addressed in magnetization conditions. 
\item Dust grain medium is in equilibrium at low power and moderate pressure. Low pressure and high power make dusty plasma highly unstable resulting in the excitation of waves \cite{mangilalmagneticdaw}. The moderate pressure (p $<$ 50 Pa) and power (P $<$ 10 W) plasma does not remain homogeneous but become inhomogeneous due to the formation of filamentary structures \cite{filamentmenati2020} in the presence of B-field. It is possible to keep plasma homogeneous (no filamentary structures) at high pressure but a much strong B-field will be required to magnetize the same-sized charged dust grains. 
\item Confinement of dust grains at a very high magnetic field ($>$ 2 Tesla) is also one of the concerns in the experiment. The confinement potential is created above the lower powered electrode (in rf discharge) by placing an additional conducting/non-conducting ring of appropriate width and thickness \cite{mangimagneticrotation}. The electric potential distribution of the power electrode as well the ring electrode is expected to change with increasing the magnetic field that can modify the confinement potential for charged dust particles at strong B-field in rf discharges.
\item The azimuthal flow of ions and electrons in the presence of dusty plasma can drive the dust particles in the same direction\cite{rotationknopka1}. The angular frequency of dust rotation depends on the strength of the magnetic field. The azimuthal flow of electrons or ions keeps dust particles in the rotational motion in the presence of B-field. It would be difficult to get a stationary equilibrium dust cloud where dust grains only experience the Lorentz force and magnetization of these massive dust particles in plasma will be happened. However, there are challenges in keeping dust grains in random motion instead of rotational motion in the presence of a strong B-field.  
\end{itemize}
%
\section{The required futuristic steps to get magnetized dusty plasma} \label{sec:secIII}
Magnetization of charged dust particles in plasma devices with the application of a very high B-field is a challenging task due to the major concerns which were discussed in the earlier section. However, more research work in the field of dusty plasma (experimental and theoretical) with a strong B-field may help to achieve the conditions of magnetization of charged particles. A few steps are suggested based on experimental findings in magnetized dusty plasma devices and published works in such devices.  
\begin{itemize}
\item As per the experimental study, negative charges on dust particles are reduced at higher magnetic field \cite{mangilaljpp,melzerdustchargeb}. It is also reported that magnetic particles can have higher charges on their surface than non-magnetic particles of the same size in magnetized plasma background \cite{mangilaljpp}. The charges on the dust surface depend on its surface area. The surface area of spherical particles increases with increasing size but same time the mass of particles also increases. To overcome this issue, hollow magnetic spherical particles can be used in place of non-magnetic solid particles. In this way, the strength of external magnetic field can be reduced to achieve the magnetization condition at moderate power and pressure. The moderate power and pressure may help in getting homogeneous plasma (without filamentary structures) at a comparatively low B-field.  
\item A large radius aluminum ring on the lower electrode in rf discharges which is used to create a potential well for charged particles provides poor confinement at a high magnetic field (B $>$ 1 T). The metal ring is in contact with the lower electrode (in rf discharge) therefore the potential (voltage) on it will be the self-biasing voltage of the lower-powered electrode. The self-bias voltage on the ring, as well as of the lower electrode, may get changed significantly due to external B-field. However, the non-conducting (Teflon/nylon) ring always remains in floating condition (floating potential) which could provide a better confinement to charged dust particles in strongly magnetized plasma. Thus, dusty plasma experiments with smaller-sized non-conducting confinement rings of appropriate thickness should be performed.
\item Experimental study of Chaubey \textit{et al.} \cite{neerajpositivechargepop,neerajpositivecrystal} in afterglow dusty plasma demonstrates the large positive charges on dust surface and holding them above the lower powered (biased) electrode without plasma background. If the levitated positively charged dust grains are exposed to the strong external magnetic field (few Tesla) then Lorentz force may act on them and start to gyrate (rotate) about B-field lines in the plane. The plasma density is very low (negligible) in afterglow discharge and charges on dust grains are frozen which may help in avoiding the issues such as the formation of filamentary structure, azimuthal flow of electrons and ions, charging mechanism of dust grains, etc. arising in normal magnetized rf discharges. However, some dedicated experiments in such after-glow discharge configuration are required to demonstrate the proposed idea for producing the magnetized dusty plasma.     
\end{itemize}
\section{Summary}  \label{sec:secIV}
In this report, we have discussed the complexity in understanding the experimental results of dusty plasma in the presence of magnetic field. The few experimentally observed results such as dust-acoustic waves, void formation, rotational motion of particles, crystallization of dust grains, etc. in magnetized dusty plasma device at Justus-Liebig University Giessen, Germany are discussed. The complexity in explaining the experimentally observed results of dusty plasma (weakly or strongly magnetized) without appropriate theoretical models/computational experiments is highlighted. The problems faced in achieving magnetization conditions like the formation of filamentary structures, poor particle confinement, and getting stationary dust grain medium at strong B-field are mentioned. Does it possible to come out from some of these challenges in achieving magnetization conditions? To get the answer of questions, a few ideas based on experimental knowledge of the subject are proposed. These suggested steps may work in achieving the magnetization conditions of charged dust particles in rf discharges at strong B-field. However, the highlighted complexity and issues in achieving magnetized dusty plasma may motivate researchers to work on these important research areas.       

\section{Acknowledgement} 
Author is thankful to Prof. Markus H Thoma for providing experimental facility at I-Institute of Physics, Justus-Liebig University Giessen, Germany.
\bibliography{Bibliography}

\begin{thebibliography}{88}%
\makeatletter
\providecommand \@ifxundefined [1]{%
 \@ifx{#1\undefined}
}%
\providecommand \@ifnum [1]{%
 \ifnum #1\expandafter \@firstoftwo
 \else \expandafter \@secondoftwo
 \fi
}%
\providecommand \@ifx [1]{%
 \ifx #1\expandafter \@firstoftwo
 \else \expandafter \@secondoftwo
 \fi
}%
\providecommand \natexlab [1]{#1}%
\providecommand \enquote  [1]{``#1''}%
\providecommand \bibnamefont  [1]{#1}%
\providecommand \bibfnamefont [1]{#1}%
\providecommand \citenamefont [1]{#1}%
\providecommand \href@noop [0]{\@secondoftwo}%
\providecommand \href [0]{\begingroup \@sanitize@url \@href}%
\providecommand \@href[1]{\@@startlink{#1}\@@href}%
\providecommand \@@href[1]{\endgroup#1\@@endlink}%
\providecommand \@sanitize@url [0]{\catcode `\\12\catcode `\$12\catcode
  `\&12\catcode `\#12\catcode `\^12\catcode `\_12\catcode `\%12\relax}%
\providecommand \@@startlink[1]{}%
\providecommand \@@endlink[0]{}%
\providecommand \url  [0]{\begingroup\@sanitize@url \@url }%
\providecommand \@url [1]{\endgroup\@href {#1}{\urlprefix }}%
\providecommand \urlprefix  [0]{URL }%
\providecommand \Eprint [0]{\href }%
\providecommand \doibase [0]{https://doi.org/}%
\providecommand \selectlanguage [0]{\@gobble}%
\providecommand \bibinfo  [0]{\@secondoftwo}%
\providecommand \bibfield  [0]{\@secondoftwo}%
\providecommand \translation [1]{[#1]}%
\providecommand \BibitemOpen [0]{}%
\providecommand \bibitemStop [0]{}%
\providecommand \bibitemNoStop [0]{.\EOS\space}%
\providecommand \EOS [0]{\spacefactor3000\relax}%
\providecommand \BibitemShut  [1]{\csname bibitem#1\endcsname}%
\let\auto@bib@innerbib\@empty
\bibitem [{\citenamefont {Chen}(1984)}]{chenplasmaphysicsbook}%
  \BibitemOpen
  \bibfield  {author} {\bibinfo {author} {\bibfnamefont {F.~F.}\ \bibnamefont
  {Chen}},\ }\href@noop {} {\emph {\bibinfo {title} {Introduction to plasma
  physics and controlled fusion}}}\ (\bibinfo  {publisher} {Springer US},\
  \bibinfo {year} {1984})\BibitemShut {NoStop}%
\bibitem [{\citenamefont {Barkan}, \citenamefont {D'Angelo},\ and\
  \citenamefont {Merlino}(1994{\natexlab{a}})}]{Charging}%
  \BibitemOpen
  \bibfield  {author} {\bibinfo {author} {\bibfnamefont {A.}~\bibnamefont
  {Barkan}}, \bibinfo {author} {\bibfnamefont {N.}~\bibnamefont {D'Angelo}},\
  and\ \bibinfo {author} {\bibfnamefont {R.~L.}\ \bibnamefont {Merlino}},\
  }\bibfield  {title} {\enquote {\bibinfo {title} {Charging of dust grains in a
  plasma},}\ }\href@noop {} {\bibfield  {journal} {\bibinfo  {journal} {Phys.
  Rev. Lett.}\ }\textbf {\bibinfo {volume} {73}},\ \bibinfo {pages}
  {3093--3096} (\bibinfo {year} {1994}{\natexlab{a}})}\BibitemShut {NoStop}%
\bibitem [{\citenamefont {Goree}(1994)}]{dustcharginggoree}%
  \BibitemOpen
  \bibfield  {author} {\bibinfo {author} {\bibfnamefont {J.}~\bibnamefont
  {Goree}},\ }\bibfield  {title} {\enquote {\bibinfo {title} {Charging of
  particles in a plasma},}\ }\href@noop {} {\bibfield  {journal} {\bibinfo
  {journal} {Plasma Sources Science and Technology}\ }\textbf {\bibinfo
  {volume} {3}},\ \bibinfo {pages} {400--406} (\bibinfo {year}
  {1994})}\BibitemShut {NoStop}%
\bibitem [{\citenamefont {Nakamura}(2002)}]{diaschokwaves}%
  \BibitemOpen
  \bibfield  {author} {\bibinfo {author} {\bibfnamefont {Y.}~\bibnamefont
  {Nakamura}},\ }\bibfield  {title} {\enquote {\bibinfo {title} {Experiments on
  ion-acoustic shock waves in a dusty plasma},}\ }\href@noop {} {\bibfield
  {journal} {\bibinfo  {journal} {Physics of Plasmas}\ }\textbf {\bibinfo
  {volume} {9}},\ \bibinfo {pages} {440--445} (\bibinfo {year}
  {2002})}\BibitemShut {NoStop}%
\bibitem [{\citenamefont {Nakamura}\ and\ \citenamefont
  {Sarma}(2001)}]{diawexp1}%
  \BibitemOpen
  \bibfield  {author} {\bibinfo {author} {\bibfnamefont {Y.}~\bibnamefont
  {Nakamura}}\ and\ \bibinfo {author} {\bibfnamefont {A.}~\bibnamefont
  {Sarma}},\ }\bibfield  {title} {\enquote {\bibinfo {title} {Observation of
  ion-acoustic solitary waves in a dusty plasma},}\ }\href@noop {} {\bibfield
  {journal} {\bibinfo  {journal} {Physics of Plasmas}\ }\textbf {\bibinfo
  {volume} {8}},\ \bibinfo {pages} {3921--3926} (\bibinfo {year}
  {2001})}\BibitemShut {NoStop}%
\bibitem [{\citenamefont {Bilik}\ \emph {et~al.}(2015)\citenamefont {Bilik},
  \citenamefont {Anthony}, \citenamefont {Merritt}, \citenamefont {Aydil},\
  and\ \citenamefont {Kortshagen}}]{eedfindusty1}%
  \BibitemOpen
  \bibfield  {author} {\bibinfo {author} {\bibfnamefont {N.}~\bibnamefont
  {Bilik}}, \bibinfo {author} {\bibfnamefont {R.}~\bibnamefont {Anthony}},
  \bibinfo {author} {\bibfnamefont {B.~A.}\ \bibnamefont {Merritt}}, \bibinfo
  {author} {\bibfnamefont {E.~S.}\ \bibnamefont {Aydil}},\ and\ \bibinfo
  {author} {\bibfnamefont {U.~R.}\ \bibnamefont {Kortshagen}},\ }\bibfield
  {title} {\enquote {\bibinfo {title} {Langmuir probe measurements of electron
  energy probability functions in dusty plasmas},}\ }\href@noop {} {\bibfield
  {journal} {\bibinfo  {journal} {Journal of Physics D: Applied Physics}\
  }\textbf {\bibinfo {volume} {48}},\ \bibinfo {pages} {105204} (\bibinfo
  {year} {2015})}\BibitemShut {NoStop}%
\bibitem [{\citenamefont {Merlino}(2014)}]{dawmerlino}%
  \BibitemOpen
  \bibfield  {author} {\bibinfo {author} {\bibfnamefont {R.~L.}\ \bibnamefont
  {Merlino}},\ }\bibfield  {title} {\enquote {\bibinfo {title} {25 years of
  dust acoustic waves},}\ }\href@noop {} {\bibfield  {journal} {\bibinfo
  {journal} {J. Plasma Physics}\ }\textbf {\bibinfo {volume} {80}},\ \bibinfo
  {pages} {773--786} (\bibinfo {year} {2014})}\BibitemShut {NoStop}%
\bibitem [{\citenamefont {D'yachkov}, \citenamefont {Petrov},\ and\
  \citenamefont {Fortov}()}]{dcdustyplasma}%
  \BibitemOpen
  \bibfield  {author} {\bibinfo {author} {\bibfnamefont {L.~G.}\ \bibnamefont
  {D'yachkov}}, \bibinfo {author} {\bibfnamefont {O.~F.}\ \bibnamefont
  {Petrov}},\ and\ \bibinfo {author} {\bibfnamefont {V.~E.}\ \bibnamefont
  {Fortov}},\ }\bibfield  {title} {\enquote {\bibinfo {title} {Dusty plasma
  structures in magnetic dc discharges},}\ }\href@noop {} {\bibfield  {journal}
  {\bibinfo  {journal} {Contributions to Plasma Physics}\ }\textbf {\bibinfo
  {volume} {49}},\ \bibinfo {pages} {134--147}}\BibitemShut {NoStop}%
\bibitem [{\citenamefont {Bandyopadhyay}\ \emph {et~al.}(2008)\citenamefont
  {Bandyopadhyay}, \citenamefont {Prasad}, \citenamefont {Sen},\ and\
  \citenamefont {Kaw}}]{pdasw}%
  \BibitemOpen
  \bibfield  {author} {\bibinfo {author} {\bibfnamefont {P.}~\bibnamefont
  {Bandyopadhyay}}, \bibinfo {author} {\bibfnamefont {G.}~\bibnamefont
  {Prasad}}, \bibinfo {author} {\bibfnamefont {A.}~\bibnamefont {Sen}},\ and\
  \bibinfo {author} {\bibfnamefont {P.~K.}\ \bibnamefont {Kaw}},\ }\bibfield
  {title} {\enquote {\bibinfo {title} {Experimental study of nonlinear dust
  acoustic solitary waves in a dusty plasma},}\ }\href@noop {} {\bibfield
  {journal} {\bibinfo  {journal} {Phys. Rev. Lett.}\ }\textbf {\bibinfo
  {volume} {101}},\ \bibinfo {pages} {065006} (\bibinfo {year}
  {2008})}\BibitemShut {NoStop}%
\bibitem [{\citenamefont {Choudhary}, \citenamefont {Mukherjee},\ and\
  \citenamefont {Bandyopadhyay}(2016{\natexlab{a}})}]{mangilalpop}%
  \BibitemOpen
  \bibfield  {author} {\bibinfo {author} {\bibfnamefont {M.}~\bibnamefont
  {Choudhary}}, \bibinfo {author} {\bibfnamefont {S.}~\bibnamefont
  {Mukherjee}},\ and\ \bibinfo {author} {\bibfnamefont {P.}~\bibnamefont
  {Bandyopadhyay}},\ }\bibfield  {title} {\enquote {\bibinfo {title}
  {Propagation characteristics of dust–acoustic waves in presence of a
  floating cylindrical object in the dc discharge plasma},}\ }\href@noop {}
  {\bibfield  {journal} {\bibinfo  {journal} {Physics of Plasmas}\ }\textbf
  {\bibinfo {volume} {23}},\ \bibinfo {pages} {083705} (\bibinfo {year}
  {2016}{\natexlab{a}})}\BibitemShut {NoStop}%
\bibitem [{\citenamefont {Choudhary}, \citenamefont {Mukherjee},\ and\
  \citenamefont {Bandyopadhyay}(2016{\natexlab{b}})}]{mangirsiexpsystem}%
  \BibitemOpen
  \bibfield  {author} {\bibinfo {author} {\bibfnamefont {M.}~\bibnamefont
  {Choudhary}}, \bibinfo {author} {\bibfnamefont {S.}~\bibnamefont
  {Mukherjee}},\ and\ \bibinfo {author} {\bibfnamefont {P.}~\bibnamefont
  {Bandyopadhyay}},\ }\bibfield  {title} {\enquote {\bibinfo {title} {Transport
  and trapping of dust particles in a potential well created by inductively
  coupled diffused plasmas},}\ }\href@noop {} {\bibfield  {journal} {\bibinfo
  {journal} {Rev. Sci. Instrum.}\ }\textbf {\bibinfo {volume} {87}},\ \bibinfo
  {pages} {053505} (\bibinfo {year} {2016}{\natexlab{b}})}\BibitemShut
  {NoStop}%
\bibitem [{\citenamefont {Chaubey}\ and\ \citenamefont
  {Goree}(2022{\natexlab{a}})}]{neerajcrystalccp1}%
  \BibitemOpen
  \bibfield  {author} {\bibinfo {author} {\bibfnamefont {N.}~\bibnamefont
  {Chaubey}}\ and\ \bibinfo {author} {\bibfnamefont {J.}~\bibnamefont
  {Goree}},\ }\bibfield  {title} {\enquote {\bibinfo {title} {Preservation of a
  dust crystal as it falls in an afterglow plasma},}\ }\href@noop {} {\bibfield
   {journal} {\bibinfo  {journal} {Frontiers in Physics}\ }\textbf {\bibinfo
  {volume} {10}} (\bibinfo {year} {2022}{\natexlab{a}})}\BibitemShut {NoStop}%
\bibitem [{\citenamefont {Schwabe}\ \emph {et~al.}(2007)\citenamefont
  {Schwabe}, \citenamefont {Rubin-Zuzic}, \citenamefont {Zhdanov},
  \citenamefont {Thomas},\ and\ \citenamefont {Morfill}}]{ddw1}%
  \BibitemOpen
  \bibfield  {author} {\bibinfo {author} {\bibfnamefont {M.}~\bibnamefont
  {Schwabe}}, \bibinfo {author} {\bibfnamefont {M.}~\bibnamefont
  {Rubin-Zuzic}}, \bibinfo {author} {\bibfnamefont {S.}~\bibnamefont
  {Zhdanov}}, \bibinfo {author} {\bibfnamefont {H.~M.}\ \bibnamefont
  {Thomas}},\ and\ \bibinfo {author} {\bibfnamefont {G.~E.}\ \bibnamefont
  {Morfill}},\ }\bibfield  {title} {\enquote {\bibinfo {title} {Highly resolved
  self-excited density waves in a complex plasma},}\ }\href@noop {} {\bibfield
  {journal} {\bibinfo  {journal} {Phys. Rev. Lett.}\ }\textbf {\bibinfo
  {volume} {99}},\ \bibinfo {pages} {095002} (\bibinfo {year}
  {2007})}\BibitemShut {NoStop}%
\bibitem [{\citenamefont {Shukla}\ and\ \citenamefont {Mamun}(2001)}]{dasw}%
  \BibitemOpen
  \bibfield  {author} {\bibinfo {author} {\bibfnamefont {P.~K.}\ \bibnamefont
  {Shukla}}\ and\ \bibinfo {author} {\bibfnamefont {A.~A.}\ \bibnamefont
  {Mamun}},\ }\bibfield  {title} {\enquote {\bibinfo {title} {Dust-acoustic
  shocks in a strongly coupled dusty plasma},}\ }\href@noop {} {\bibfield
  {journal} {\bibinfo  {journal} {IEEE Trans. Plasma Sci.}\ }\textbf {\bibinfo
  {volume} {29}},\ \bibinfo {pages} {221--225} (\bibinfo {year}
  {2001})}\BibitemShut {NoStop}%
\bibitem [{\citenamefont {Thompson}\ \emph {et~al.}(1997)\citenamefont
  {Thompson}, \citenamefont {Barkan}, \citenamefont {D'Angelo},\ and\
  \citenamefont {Merlino}}]{daw3}%
  \BibitemOpen
  \bibfield  {author} {\bibinfo {author} {\bibfnamefont {C.}~\bibnamefont
  {Thompson}}, \bibinfo {author} {\bibfnamefont {A.}~\bibnamefont {Barkan}},
  \bibinfo {author} {\bibfnamefont {N.}~\bibnamefont {D'Angelo}},\ and\
  \bibinfo {author} {\bibfnamefont {R.~L.}\ \bibnamefont {Merlino}},\
  }\bibfield  {title} {\enquote {\bibinfo {title} {Dust acoustic waves in a
  direct current glow discharge},}\ }\href@noop {} {\bibfield  {journal}
  {\bibinfo  {journal} {Phys. Plasmas}\ }\textbf {\bibinfo {volume} {4}},\
  \bibinfo {pages} {2331--2335} (\bibinfo {year} {1997})}\BibitemShut {NoStop}%
\bibitem [{\citenamefont {Flanagan}\ and\ \citenamefont {Goree}(2010)}]{ddw2}%
  \BibitemOpen
  \bibfield  {author} {\bibinfo {author} {\bibfnamefont {T.~M.}\ \bibnamefont
  {Flanagan}}\ and\ \bibinfo {author} {\bibfnamefont {J.}~\bibnamefont
  {Goree}},\ }\bibfield  {title} {\enquote {\bibinfo {title} {Observation of
  the spatial growth of self-excited dust-density waves},}\ }\href@noop {}
  {\bibfield  {journal} {\bibinfo  {journal} {Phys. Plasmas}\ }\textbf
  {\bibinfo {volume} {17}},\ \bibinfo {pages} {123702} (\bibinfo {year}
  {2010})}\BibitemShut {NoStop}%
\bibitem [{\citenamefont {Molotkov}\ \emph {et~al.}(2004)\citenamefont
  {Molotkov}, \citenamefont {Petrov}, \citenamefont {Pustyl’nik},
  \citenamefont {Torchinskii}, \citenamefont {Fortov},\ and\ \citenamefont
  {Khrapak}}]{ddwfortov}%
  \BibitemOpen
  \bibfield  {author} {\bibinfo {author} {\bibfnamefont {V.~I.}\ \bibnamefont
  {Molotkov}}, \bibinfo {author} {\bibfnamefont {O.~F.}\ \bibnamefont
  {Petrov}}, \bibinfo {author} {\bibfnamefont {M.~Y.}\ \bibnamefont
  {Pustyl’nik}}, \bibinfo {author} {\bibfnamefont {V.~M.}\ \bibnamefont
  {Torchinskii}}, \bibinfo {author} {\bibfnamefont {V.~E.}\ \bibnamefont
  {Fortov}},\ and\ \bibinfo {author} {\bibfnamefont {A.~G.}\ \bibnamefont
  {Khrapak}},\ }\bibfield  {title} {\enquote {\bibinfo {title} {Dusty plasma of
  a dc glow discharge: methods of investigation and characteristic features of
  behavior},}\ }\href@noop {} {\bibfield  {journal} {\bibinfo  {journal} {High
  Temperature}\ }\textbf {\bibinfo {volume} {42}},\ \bibinfo {pages} {827--841}
  (\bibinfo {year} {2004})}\BibitemShut {NoStop}%
\bibitem [{\citenamefont {Karasev}\ \emph
  {et~al.}(2016{\natexlab{a}})\citenamefont {Karasev}, \citenamefont {Dzlieva},
  \citenamefont {Pavlov}, \citenamefont {Ermolenko}, \citenamefont {Novikov},\
  and\ \citenamefont {Maiorov}}]{rotationkarasev2}%
  \BibitemOpen
  \bibfield  {author} {\bibinfo {author} {\bibfnamefont {V.~Y.}\ \bibnamefont
  {Karasev}}, \bibinfo {author} {\bibfnamefont {E.~S.}\ \bibnamefont
  {Dzlieva}}, \bibinfo {author} {\bibfnamefont {S.~I.}\ \bibnamefont {Pavlov}},
  \bibinfo {author} {\bibfnamefont {M.~A.}\ \bibnamefont {Ermolenko}}, \bibinfo
  {author} {\bibfnamefont {L.~A.}\ \bibnamefont {Novikov}},\ and\ \bibinfo
  {author} {\bibfnamefont {S.~A.}\ \bibnamefont {Maiorov}},\ }\bibfield
  {title} {\enquote {\bibinfo {title} {The dynamics of dust structures under
  magnetic field in stratified glow discharge},}\ }\href@noop {} {\bibfield
  {journal} {\bibinfo  {journal} {Contributions to Plasma Physics}\ }\textbf
  {\bibinfo {volume} {56}},\ \bibinfo {pages} {197--203} (\bibinfo {year}
  {2016}{\natexlab{a}})}\BibitemShut {NoStop}%
\bibitem [{\citenamefont {Choudhary}\ \emph {et~al.}(2021)\citenamefont
  {Choudhary}, \citenamefont {Bergert}, \citenamefont {Moritz}, \citenamefont
  {Mitic},\ and\ \citenamefont {Thoma}}]{mangilalannulardusty}%
  \BibitemOpen
  \bibfield  {author} {\bibinfo {author} {\bibfnamefont {M.}~\bibnamefont
  {Choudhary}}, \bibinfo {author} {\bibfnamefont {R.}~\bibnamefont {Bergert}},
  \bibinfo {author} {\bibfnamefont {S.}~\bibnamefont {Moritz}}, \bibinfo
  {author} {\bibfnamefont {S.}~\bibnamefont {Mitic}},\ and\ \bibinfo {author}
  {\bibfnamefont {M.~H.}\ \bibnamefont {Thoma}},\ }\bibfield  {title} {\enquote
  {\bibinfo {title} {Rotational properties of annulus dusty plasma in a strong
  magnetic field},}\ }\href@noop {} {\bibfield  {journal} {\bibinfo  {journal}
  {Contributions to Plasma Physics}\ }\textbf {\bibinfo {volume} {61}},\
  \bibinfo {pages} {e202000110} (\bibinfo {year} {2021})}\BibitemShut {NoStop}%
\bibitem [{\citenamefont {Bockwoldt}\ \emph {et~al.}(2014)\citenamefont
  {Bockwoldt}, \citenamefont {Arp}, \citenamefont {Menzel},\ and\ \citenamefont
  {Piel}}]{vortexmicrogravity}%
  \BibitemOpen
  \bibfield  {author} {\bibinfo {author} {\bibfnamefont {T.}~\bibnamefont
  {Bockwoldt}}, \bibinfo {author} {\bibfnamefont {O.}~\bibnamefont {Arp}},
  \bibinfo {author} {\bibfnamefont {K.~O.}\ \bibnamefont {Menzel}},\ and\
  \bibinfo {author} {\bibfnamefont {A.}~\bibnamefont {Piel}},\ }\bibfield
  {title} {\enquote {\bibinfo {title} {On the origin of dust vortices in
  complex plasmas under microgravity conditions},}\ }\href@noop {} {\bibfield
  {journal} {\bibinfo  {journal} {Phys. Plasmas}\ }\textbf {\bibinfo {volume}
  {21}},\ \bibinfo {pages} {103703} (\bibinfo {year} {2014})}\BibitemShut
  {NoStop}%
\bibitem [{\citenamefont {Choudhary}, \citenamefont {Mukherjee},\ and\
  \citenamefont {Bandyopadhyay}(2017)}]{mangilalmultiplerot}%
  \BibitemOpen
  \bibfield  {author} {\bibinfo {author} {\bibfnamefont {M.}~\bibnamefont
  {Choudhary}}, \bibinfo {author} {\bibfnamefont {S.}~\bibnamefont
  {Mukherjee}},\ and\ \bibinfo {author} {\bibfnamefont {P.}~\bibnamefont
  {Bandyopadhyay}},\ }\bibfield  {title} {\enquote {\bibinfo {title}
  {Experimental observation of self excited co-rotating multiple vortices in a
  dusty plasma with inhomogeneous plasma background},}\ }\href@noop {}
  {\bibfield  {journal} {\bibinfo  {journal} {Physics of Plasmas}\ }\textbf
  {\bibinfo {volume} {24}},\ \bibinfo {pages} {033703} (\bibinfo {year}
  {2017})}\BibitemShut {NoStop}%
\bibitem [{\citenamefont {Singh~Dharodi}, \citenamefont {Kumar~Tiwari},\ and\
  \citenamefont {Das}(2014)}]{vikramtsw}%
  \BibitemOpen
  \bibfield  {author} {\bibinfo {author} {\bibfnamefont {V.}~\bibnamefont
  {Singh~Dharodi}}, \bibinfo {author} {\bibfnamefont {S.}~\bibnamefont
  {Kumar~Tiwari}},\ and\ \bibinfo {author} {\bibfnamefont {A.}~\bibnamefont
  {Das}},\ }\bibfield  {title} {\enquote {\bibinfo {title} {Visco-elastic fluid
  simulations of coherent structures in strongly coupled dusty plasma
  medium},}\ }\href@noop {} {\bibfield  {journal} {\bibinfo  {journal} {Physics
  of Plasmas}\ }\textbf {\bibinfo {volume} {21}},\ \bibinfo {pages} {073705}
  (\bibinfo {year} {2014})}\BibitemShut {NoStop}%
\bibitem [{\citenamefont {Choudhary}, \citenamefont {Mukherjee},\ and\
  \citenamefont {Bandyopadhyay}(2018)}]{mangilargeaspect}%
  \BibitemOpen
  \bibfield  {author} {\bibinfo {author} {\bibfnamefont {M.}~\bibnamefont
  {Choudhary}}, \bibinfo {author} {\bibfnamefont {S.}~\bibnamefont
  {Mukherjee}},\ and\ \bibinfo {author} {\bibfnamefont {P.}~\bibnamefont
  {Bandyopadhyay}},\ }\bibfield  {title} {\enquote {\bibinfo {title}
  {Collective dynamics of large aspect ratio dusty plasma in an inhomogeneous
  plasma background: Formation of the co-rotating vortex series},}\ }\href@noop
  {} {\bibfield  {journal} {\bibinfo  {journal} {Physics of Plasmas}\ }\textbf
  {\bibinfo {volume} {25}},\ \bibinfo {pages} {023704} (\bibinfo {year}
  {2018})}\BibitemShut {NoStop}%
\bibitem [{\citenamefont {Laishram}\ \emph {et~al.}(2017)\citenamefont
  {Laishram}, \citenamefont {Sharma}, \citenamefont {Chattopdhyay},\ and\
  \citenamefont {Kaw}}]{modhuvortices}%
  \BibitemOpen
  \bibfield  {author} {\bibinfo {author} {\bibfnamefont {M.}~\bibnamefont
  {Laishram}}, \bibinfo {author} {\bibfnamefont {D.}~\bibnamefont {Sharma}},
  \bibinfo {author} {\bibfnamefont {P.~K.}\ \bibnamefont {Chattopdhyay}},\ and\
  \bibinfo {author} {\bibfnamefont {P.~K.}\ \bibnamefont {Kaw}},\ }\bibfield
  {title} {\enquote {\bibinfo {title} {Nonlinear effects in the bounded
  dust-vortex flow in plasma},}\ }\href@noop {} {\bibfield  {journal} {\bibinfo
   {journal} {Phys. Rev. E}\ }\textbf {\bibinfo {volume} {95}},\ \bibinfo
  {pages} {033204} (\bibinfo {year} {2017})}\BibitemShut {NoStop}%
\bibitem [{\citenamefont {Chai}\ and\ \citenamefont
  {Bellan}(2016)}]{bellanicedustyrotation}%
  \BibitemOpen
  \bibfield  {author} {\bibinfo {author} {\bibfnamefont {K.-B.}\ \bibnamefont
  {Chai}}\ and\ \bibinfo {author} {\bibfnamefont {P.~M.}\ \bibnamefont
  {Bellan}},\ }\bibfield  {title} {\enquote {\bibinfo {title} {Vortex motion of
  dust particles due to non-conservative ion drag force in a plasma},}\
  }\href@noop {} {\bibfield  {journal} {\bibinfo  {journal} {Phys. Plasmas}\
  }\textbf {\bibinfo {volume} {23}},\ \bibinfo {pages} {023701} (\bibinfo
  {year} {2016})}\BibitemShut {NoStop}%
\bibitem [{\citenamefont {Thomas}\ \emph
  {et~al.}(1994{\natexlab{a}})\citenamefont {Thomas}, \citenamefont {Morfill},
  \citenamefont {Demmel}, \citenamefont {Goree}, \citenamefont {Feuerbacher},\
  and\ \citenamefont {M\"ohlmann}}]{thomasdustycrystal}%
  \BibitemOpen
  \bibfield  {author} {\bibinfo {author} {\bibfnamefont {H.}~\bibnamefont
  {Thomas}}, \bibinfo {author} {\bibfnamefont {G.~E.}\ \bibnamefont {Morfill}},
  \bibinfo {author} {\bibfnamefont {V.}~\bibnamefont {Demmel}}, \bibinfo
  {author} {\bibfnamefont {J.}~\bibnamefont {Goree}}, \bibinfo {author}
  {\bibfnamefont {B.}~\bibnamefont {Feuerbacher}},\ and\ \bibinfo {author}
  {\bibfnamefont {D.}~\bibnamefont {M\"ohlmann}},\ }\bibfield  {title}
  {\enquote {\bibinfo {title} {Plasma crystal: Coulomb crystallization in a
  dusty plasma},}\ }\href@noop {} {\bibfield  {journal} {\bibinfo  {journal}
  {Phys. Rev. Lett.}\ }\textbf {\bibinfo {volume} {73}},\ \bibinfo {pages}
  {652--655} (\bibinfo {year} {1994}{\natexlab{a}})}\BibitemShut {NoStop}%
\bibitem [{\citenamefont {Chu}\ and\ \citenamefont
  {I}(1994)}]{linidustycrystal2}%
  \BibitemOpen
  \bibfield  {author} {\bibinfo {author} {\bibfnamefont {J.~H.}\ \bibnamefont
  {Chu}}\ and\ \bibinfo {author} {\bibfnamefont {L.}~\bibnamefont {I}},\
  }\bibfield  {title} {\enquote {\bibinfo {title} {Direct observation of
  coulomb crystals and liquids in strongly coupled rf dusty plasmas},}\
  }\href@noop {} {\bibfield  {journal} {\bibinfo  {journal} {Phys. Rev. Lett.}\
  }\textbf {\bibinfo {volume} {72}},\ \bibinfo {pages} {4009--4012} (\bibinfo
  {year} {1994})}\BibitemShut {NoStop}%
\bibitem [{\citenamefont {Hariprasad}\ \emph {et~al.}(2018)\citenamefont
  {Hariprasad}, \citenamefont {Bandyopadhyay}, \citenamefont {Arora},\ and\
  \citenamefont {Sen}}]{dustcrystalhariparasad}%
  \BibitemOpen
  \bibfield  {author} {\bibinfo {author} {\bibfnamefont {M.~G.}\ \bibnamefont
  {Hariprasad}}, \bibinfo {author} {\bibfnamefont {P.}~\bibnamefont
  {Bandyopadhyay}}, \bibinfo {author} {\bibfnamefont {G.}~\bibnamefont
  {Arora}},\ and\ \bibinfo {author} {\bibfnamefont {A.}~\bibnamefont {Sen}},\
  }\bibfield  {title} {\enquote {\bibinfo {title} {{Experimental observation of
  a dusty plasma crystal in the cathode sheath of a DC glow discharge
  plasma}},}\ }\href@noop {} {\bibfield  {journal} {\bibinfo  {journal}
  {Physics of Plasmas}\ }\textbf {\bibinfo {volume} {25}},\ \bibinfo {pages}
  {123704} (\bibinfo {year} {2018})}\BibitemShut {NoStop}%
\bibitem [{\citenamefont {Melzer}, \citenamefont {Homann},\ and\ \citenamefont
  {Piel}(1996)}]{melzerdustcrystalmelting}%
  \BibitemOpen
  \bibfield  {author} {\bibinfo {author} {\bibfnamefont {A.}~\bibnamefont
  {Melzer}}, \bibinfo {author} {\bibfnamefont {A.}~\bibnamefont {Homann}},\
  and\ \bibinfo {author} {\bibfnamefont {A.}~\bibnamefont {Piel}},\ }\bibfield
  {title} {\enquote {\bibinfo {title} {Experimental investigation of the
  melting transition of the plasma crystal},}\ }\href@noop {} {\bibfield
  {journal} {\bibinfo  {journal} {Phys. Rev. E}\ }\textbf {\bibinfo {volume}
  {53}},\ \bibinfo {pages} {2757--2766} (\bibinfo {year} {1996})}\BibitemShut
  {NoStop}%
\bibitem [{\citenamefont {Tiwari}\ \emph {et~al.}(2012)\citenamefont {Tiwari},
  \citenamefont {Das}, \citenamefont {Angom}, \citenamefont {Patel},\ and\
  \citenamefont {Kaw}}]{sanatkhinstability}%
  \BibitemOpen
  \bibfield  {author} {\bibinfo {author} {\bibfnamefont {S.~K.}\ \bibnamefont
  {Tiwari}}, \bibinfo {author} {\bibfnamefont {A.}~\bibnamefont {Das}},
  \bibinfo {author} {\bibfnamefont {D.}~\bibnamefont {Angom}}, \bibinfo
  {author} {\bibfnamefont {B.~G.}\ \bibnamefont {Patel}},\ and\ \bibinfo
  {author} {\bibfnamefont {P.}~\bibnamefont {Kaw}},\ }\bibfield  {title}
  {\enquote {\bibinfo {title} {Kelvin-helmholtz instability in a strongly
  coupled dusty plasma medium},}\ }\href@noop {} {\bibfield  {journal}
  {\bibinfo  {journal} {Physics of Plasmas}\ }\textbf {\bibinfo {volume}
  {19}},\ \bibinfo {pages} {073703} (\bibinfo {year} {2012})}\BibitemShut
  {NoStop}%
\bibitem [{\citenamefont {Dharodi}, \citenamefont {Patel},\ and\ \citenamefont
  {Das}(2022)}]{vikramkhinstability}%
  \BibitemOpen
  \bibfield  {author} {\bibinfo {author} {\bibfnamefont {V.~S.}\ \bibnamefont
  {Dharodi}}, \bibinfo {author} {\bibfnamefont {B.}~\bibnamefont {Patel}},\
  and\ \bibinfo {author} {\bibfnamefont {A.}~\bibnamefont {Das}},\ }\bibfield
  {title} {\enquote {\bibinfo {title} {Kelvin–helmholtz instability in
  strongly coupled dusty plasma with rotational shear flows and tracer
  transport},}\ }\href@noop {} {\bibfield  {journal} {\bibinfo  {journal}
  {Journal of Plasma Physics}\ }\textbf {\bibinfo {volume} {88}},\ \bibinfo
  {pages} {905880103} (\bibinfo {year} {2022})}\BibitemShut {NoStop}%
\bibitem [{\citenamefont {Veeresha}, \citenamefont {Das},\ and\ \citenamefont
  {Sen}(2005)}]{rtinstabilityvortices}%
  \BibitemOpen
  \bibfield  {author} {\bibinfo {author} {\bibfnamefont {B.~M.}\ \bibnamefont
  {Veeresha}}, \bibinfo {author} {\bibfnamefont {A.}~\bibnamefont {Das}},\ and\
  \bibinfo {author} {\bibfnamefont {A.}~\bibnamefont {Sen}},\ }\bibfield
  {title} {\enquote {\bibinfo {title} {Rayleigh--taylor instability driven
  nonlinear vortices in dusty plasmas},}\ }\href@noop {} {\bibfield  {journal}
  {\bibinfo  {journal} {Phys. Plasmas}\ }\textbf {\bibinfo {volume} {12}},\
  \bibinfo {pages} {044506} (\bibinfo {year} {2005})}\BibitemShut {NoStop}%
\bibitem [{\citenamefont {Avinash}\ and\ \citenamefont
  {Sen}(2015)}]{avinashrtinstability}%
  \BibitemOpen
  \bibfield  {author} {\bibinfo {author} {\bibfnamefont {K.}~\bibnamefont
  {Avinash}}\ and\ \bibinfo {author} {\bibfnamefont {A.}~\bibnamefont {Sen}},\
  }\bibfield  {title} {\enquote {\bibinfo {title} {Rayleigh-taylor instability
  in dusty plasma experiment},}\ }\href@noop {} {\bibfield  {journal} {\bibinfo
   {journal} {Physics of Plasmas}\ }\textbf {\bibinfo {volume} {22}},\ \bibinfo
  {pages} {083707} (\bibinfo {year} {2015})}\BibitemShut {NoStop}%
\bibitem [{\citenamefont {Dharodi}\ and\ \citenamefont
  {Das}()}]{vikramrtinstability}%
  \BibitemOpen
  \bibfield  {author} {\bibinfo {author} {\bibfnamefont {V.~S.}\ \bibnamefont
  {Dharodi}}\ and\ \bibinfo {author} {\bibfnamefont {A.}~\bibnamefont {Das}},\
  }\href@noop {} {\ }\BibitemShut {NoStop}%
\bibitem [{\citenamefont {Thomas}\ \emph {et~al.}(2004)\citenamefont {Thomas},
  \citenamefont {Jr.}, \citenamefont {Avinash},\ and\ \citenamefont
  {Merlino}}]{void}%
  \BibitemOpen
  \bibfield  {author} {\bibinfo {author} {\bibfnamefont {E.}~\bibnamefont
  {Thomas}}, \bibinfo {author} {\bibnamefont {Jr.}}, \bibinfo {author}
  {\bibfnamefont {K.}~\bibnamefont {Avinash}},\ and\ \bibinfo {author}
  {\bibfnamefont {R.~L.}\ \bibnamefont {Merlino}},\ }\bibfield  {title}
  {\enquote {\bibinfo {title} {Probe induced voids in a dusty plasma},}\
  }\href@noop {} {\bibfield  {journal} {\bibinfo  {journal} {Phys. Plasmas}\
  }\textbf {\bibinfo {volume} {11}},\ \bibinfo {pages} {1770--1774} (\bibinfo
  {year} {2004})}\BibitemShut {NoStop}%
\bibitem [{\citenamefont {Harris}, \citenamefont {Matthews},\ and\
  \citenamefont {Hyde}(2015)}]{cavity}%
  \BibitemOpen
  \bibfield  {author} {\bibinfo {author} {\bibfnamefont {B.~J.}\ \bibnamefont
  {Harris}}, \bibinfo {author} {\bibfnamefont {L.~S.}\ \bibnamefont
  {Matthews}},\ and\ \bibinfo {author} {\bibfnamefont {T.~W.}\ \bibnamefont
  {Hyde}},\ }\bibfield  {title} {\enquote {\bibinfo {title} {Dusty plasma
  cavities: Probe-induced and natural},}\ }\href@noop {} {\bibfield  {journal}
  {\bibinfo  {journal} {Phys. Rev. E}\ }\textbf {\bibinfo {volume} {91}},\
  \bibinfo {pages} {063105} (\bibinfo {year} {2015})}\BibitemShut {NoStop}%
\bibitem [{\citenamefont {Sarkar}\ \emph {et~al.}(2015)\citenamefont {Sarkar},
  \citenamefont {Mondal}, \citenamefont {Bose},\ and\ \citenamefont
  {Mukherjee}}]{ringvoid}%
  \BibitemOpen
  \bibfield  {author} {\bibinfo {author} {\bibfnamefont {S.}~\bibnamefont
  {Sarkar}}, \bibinfo {author} {\bibfnamefont {M.}~\bibnamefont {Mondal}},
  \bibinfo {author} {\bibfnamefont {M.}~\bibnamefont {Bose}},\ and\ \bibinfo
  {author} {\bibfnamefont {S.}~\bibnamefont {Mukherjee}},\ }\bibfield  {title}
  {\enquote {\bibinfo {title} {Observation of external control and formation of
  a void in cogenerated dusty plasma},}\ }\href@noop {} {\bibfield  {journal}
  {\bibinfo  {journal} {Plasma Sources Sci. Technol.}\ }\textbf {\bibinfo
  {volume} {24}},\ \bibinfo {pages} {035007} (\bibinfo {year}
  {2015})}\BibitemShut {NoStop}%
\bibitem [{\citenamefont {Abdirakhmanov}, \citenamefont {Kodanova},\ and\
  \citenamefont {Ramazanov}(2022)}]{weaklymagnetized1}%
  \BibitemOpen
  \bibfield  {author} {\bibinfo {author} {\bibfnamefont {A.~R.}\ \bibnamefont
  {Abdirakhmanov}}, \bibinfo {author} {\bibfnamefont {S.~K.}\ \bibnamefont
  {Kodanova}},\ and\ \bibinfo {author} {\bibfnamefont {T.~S.}\ \bibnamefont
  {Ramazanov}},\ }\bibfield  {title} {\enquote {\bibinfo {title} {Dynamics of
  dust particles in the glow discharge stratum in crossed e $\times$b
  field},}\ }\href@noop {} {\bibfield  {journal} {\bibinfo  {journal}
  {Contributions to Plasma Physics}\ }\textbf {\bibinfo {volume} {n/a}},\
  \bibinfo {pages} {e202200148} (\bibinfo {year} {2022})}\BibitemShut {NoStop}%
\bibitem [{\citenamefont {Karasev}\ \emph
  {et~al.}(2016{\natexlab{b}})\citenamefont {Karasev}, \citenamefont {Dzlieva},
  \citenamefont {Pavlov}, \citenamefont {Ermolenko}, \citenamefont {Novikov},\
  and\ \citenamefont {Maiorov}}]{weaklymagnetized2}%
  \BibitemOpen
  \bibfield  {author} {\bibinfo {author} {\bibfnamefont {V.~Y.}\ \bibnamefont
  {Karasev}}, \bibinfo {author} {\bibfnamefont {E.~S.}\ \bibnamefont
  {Dzlieva}}, \bibinfo {author} {\bibfnamefont {S.~I.}\ \bibnamefont {Pavlov}},
  \bibinfo {author} {\bibfnamefont {M.~A.}\ \bibnamefont {Ermolenko}}, \bibinfo
  {author} {\bibfnamefont {L.~A.}\ \bibnamefont {Novikov}},\ and\ \bibinfo
  {author} {\bibfnamefont {S.~A.}\ \bibnamefont {Maiorov}},\ }\bibfield
  {title} {\enquote {\bibinfo {title} {The dynamics of dust structures under
  magnetic field in stratified glow discharge},}\ }\href@noop {} {\bibfield
  {journal} {\bibinfo  {journal} {Contributions to Plasma Physics}\ }\textbf
  {\bibinfo {volume} {56}},\ \bibinfo {pages} {197--203} (\bibinfo {year}
  {2016}{\natexlab{b}})}\BibitemShut {NoStop}%
\bibitem [{\citenamefont {Trottenberg}, \citenamefont {Block},\ and\
  \citenamefont {Piel}(2006)}]{weaklymagnetized3}%
  \BibitemOpen
  \bibfield  {author} {\bibinfo {author} {\bibfnamefont {T.}~\bibnamefont
  {Trottenberg}}, \bibinfo {author} {\bibfnamefont {D.}~\bibnamefont {Block}},\
  and\ \bibinfo {author} {\bibfnamefont {A.}~\bibnamefont {Piel}},\ }\bibfield
  {title} {\enquote {\bibinfo {title} {Dust confinement and dust-acoustic waves
  in weakly magnetized anodic plasmas},}\ }\href@noop {} {\bibfield  {journal}
  {\bibinfo  {journal} {Phys. Plasmas}\ }\textbf {\bibinfo {volume} {13}},\
  \bibinfo {pages} {042105} (\bibinfo {year} {2006})}\BibitemShut {NoStop}%
\bibitem [{\citenamefont {Dzlieva}\ \emph {et~al.}(2023)\citenamefont
  {Dzlieva}, \citenamefont {Dyachkov}, \citenamefont {Karasev}, \citenamefont
  {Novikov},\ and\ \citenamefont {Pavlov}}]{stronglymagnetizeddc1}%
  \BibitemOpen
  \bibfield  {author} {\bibinfo {author} {\bibfnamefont {E.~S.}\ \bibnamefont
  {Dzlieva}}, \bibinfo {author} {\bibfnamefont {L.~G.}\ \bibnamefont
  {Dyachkov}}, \bibinfo {author} {\bibfnamefont {V.~Y.}\ \bibnamefont
  {Karasev}}, \bibinfo {author} {\bibfnamefont {L.~A.}\ \bibnamefont
  {Novikov}},\ and\ \bibinfo {author} {\bibfnamefont {S.~I.}\ \bibnamefont
  {Pavlov}},\ }\bibfield  {title} {\enquote {\bibinfo {title} {Dusty plasma
  under conditions of glow discharge in magnetic field of up to 2.5 t},}\
  }\href@noop {} {\bibfield  {journal} {\bibinfo  {journal} {Plasma Physics
  Reports}\ }\textbf {\bibinfo {volume} {49}},\ \bibinfo {pages} {10--14}
  (\bibinfo {year} {2023})}\BibitemShut {NoStop}%
\bibitem [{\citenamefont {Thomas}\ \emph {et~al.}(2015)\citenamefont {Thomas},
  \citenamefont {Konopka}, \citenamefont {Artis}, \citenamefont {Lynch},
  \citenamefont {Leblanc}, \citenamefont {Adams}, \citenamefont {Merlino},\
  and\ \citenamefont {Rosenberg}}]{thomasmpedx}%
  \BibitemOpen
  \bibfield  {author} {\bibinfo {author} {\bibfnamefont {E.}~\bibnamefont
  {Thomas}}, \bibinfo {author} {\bibfnamefont {U.}~\bibnamefont {Konopka}},
  \bibinfo {author} {\bibfnamefont {D.}~\bibnamefont {Artis}}, \bibinfo
  {author} {\bibfnamefont {B.}~\bibnamefont {Lynch}}, \bibinfo {author}
  {\bibfnamefont {S.}~\bibnamefont {Leblanc}}, \bibinfo {author} {\bibfnamefont
  {S.}~\bibnamefont {Adams}}, \bibinfo {author} {\bibfnamefont {R.~L.}\
  \bibnamefont {Merlino}},\ and\ \bibinfo {author} {\bibfnamefont
  {M.}~\bibnamefont {Rosenberg}},\ }\bibfield  {title} {\enquote {\bibinfo
  {title} {The magnetized dusty plasma experiment (mdpx)},}\ }\href@noop {}
  {\bibfield  {journal} {\bibinfo  {journal} {Journal of Plasma Physics}\
  }\textbf {\bibinfo {volume} {81}} (\bibinfo {year} {2015})}\BibitemShut
  {NoStop}%
\bibitem [{\citenamefont {Tadsen}, \citenamefont {Greiner},\ and\ \citenamefont
  {Piel}(2018)}]{melzermagnetizeddusty}%
  \BibitemOpen
  \bibfield  {author} {\bibinfo {author} {\bibfnamefont {B.}~\bibnamefont
  {Tadsen}}, \bibinfo {author} {\bibfnamefont {F.}~\bibnamefont {Greiner}},\
  and\ \bibinfo {author} {\bibfnamefont {A.}~\bibnamefont {Piel}},\ }\bibfield
  {title} {\enquote {\bibinfo {title} {Probing a dusty magnetized plasma with
  self-excited dust-density waves},}\ }\href@noop {} {\bibfield  {journal}
  {\bibinfo  {journal} {Phys. Rev. E}\ }\textbf {\bibinfo {volume} {97}},\
  \bibinfo {pages} {033203} (\bibinfo {year} {2018})}\BibitemShut {NoStop}%
\bibitem [{\citenamefont {Choudhary}\ \emph
  {et~al.}(2020{\natexlab{a}})\citenamefont {Choudhary}, \citenamefont
  {Bergert}, \citenamefont {Mitic},\ and\ \citenamefont {Thoma}}]{mangilaljpp}%
  \BibitemOpen
  \bibfield  {author} {\bibinfo {author} {\bibfnamefont {M.}~\bibnamefont
  {Choudhary}}, \bibinfo {author} {\bibfnamefont {R.}~\bibnamefont {Bergert}},
  \bibinfo {author} {\bibfnamefont {S.}~\bibnamefont {Mitic}},\ and\ \bibinfo
  {author} {\bibfnamefont {M.~H.}\ \bibnamefont {Thoma}},\ }\bibfield  {title}
  {\enquote {\bibinfo {title} {Comparative study of the surface potential of
  magnetic and non-magnetic spherical objects in a magnetized radio-frequency
  discharge},}\ }\href@noop {} {\bibfield  {journal} {\bibinfo  {journal}
  {Journal of Plasma Physics}\ }\textbf {\bibinfo {volume} {86}},\ \bibinfo
  {pages} {905860508} (\bibinfo {year} {2020}{\natexlab{a}})}\BibitemShut
  {NoStop}%
\bibitem [{\citenamefont {Barkan}, \citenamefont {D'Angelo},\ and\
  \citenamefont {Merlino}(1994{\natexlab{b}})}]{chargingbarken}%
  \BibitemOpen
  \bibfield  {author} {\bibinfo {author} {\bibfnamefont {A.}~\bibnamefont
  {Barkan}}, \bibinfo {author} {\bibfnamefont {N.}~\bibnamefont {D'Angelo}},\
  and\ \bibinfo {author} {\bibfnamefont {R.~L.}\ \bibnamefont {Merlino}},\
  }\bibfield  {title} {\enquote {\bibinfo {title} {Charging of dust grains in a
  plasma},}\ }\href@noop {} {\bibfield  {journal} {\bibinfo  {journal} {Phys.
  Rev. Lett.}\ }\textbf {\bibinfo {volume} {73}},\ \bibinfo {pages}
  {3093--3096} (\bibinfo {year} {1994}{\natexlab{b}})}\BibitemShut {NoStop}%
\bibitem [{\citenamefont {Shukla}(2001)}]{shukladusty3}%
  \BibitemOpen
  \bibfield  {author} {\bibinfo {author} {\bibfnamefont {P.~K.}\ \bibnamefont
  {Shukla}},\ }\bibfield  {title} {\enquote {\bibinfo {title} {A survey of
  dusty plasma physics},}\ }\href@noop {} {\bibfield  {journal} {\bibinfo
  {journal} {Physics of Plasmas}\ }\textbf {\bibinfo {volume} {8}},\ \bibinfo
  {pages} {1791--1803} (\bibinfo {year} {2001})}\BibitemShut {NoStop}%
\bibitem [{\citenamefont {Tsytovich}, \citenamefont {Sato},\ and\ \citenamefont
  {Morfill}(2003)}]{tsytovichdustcharge1}%
  \BibitemOpen
  \bibfield  {author} {\bibinfo {author} {\bibfnamefont {V.~N.}\ \bibnamefont
  {Tsytovich}}, \bibinfo {author} {\bibfnamefont {N.}~\bibnamefont {Sato}},\
  and\ \bibinfo {author} {\bibfnamefont {G.~E.}\ \bibnamefont {Morfill}},\
  }\bibfield  {title} {\enquote {\bibinfo {title} {Note on the charging and
  spinning of dust particles in complex plasmas in a strong magnetic field},}\
  }\href@noop {} {\bibfield  {journal} {\bibinfo  {journal} {New Journal of
  Physics}\ }\textbf {\bibinfo {volume} {5}},\ \bibinfo {pages} {43--43}
  (\bibinfo {year} {2003})}\BibitemShut {NoStop}%
\bibitem [{\citenamefont {Melzer}\ \emph {et~al.}(2019)\citenamefont {Melzer},
  \citenamefont {Krüger}, \citenamefont {Schütt},\ and\ \citenamefont
  {Mulsow}}]{melzerdustchargeb}%
  \BibitemOpen
  \bibfield  {author} {\bibinfo {author} {\bibfnamefont {A.}~\bibnamefont
  {Melzer}}, \bibinfo {author} {\bibfnamefont {H.}~\bibnamefont {Krüger}},
  \bibinfo {author} {\bibfnamefont {S.}~\bibnamefont {Schütt}},\ and\ \bibinfo
  {author} {\bibfnamefont {M.}~\bibnamefont {Mulsow}},\ }\bibfield  {title}
  {\enquote {\bibinfo {title} {Finite dust clusters under strong magnetic
  fields},}\ }\href@noop {} {\bibfield  {journal} {\bibinfo  {journal} {Physics
  of Plasmas}\ }\textbf {\bibinfo {volume} {26}},\ \bibinfo {pages} {093702}
  (\bibinfo {year} {2019})}\BibitemShut {NoStop}%
\bibitem [{\citenamefont {Lange}(2016)}]{langefloatinginmagnetized}%
  \BibitemOpen
  \bibfield  {author} {\bibinfo {author} {\bibfnamefont {D.}~\bibnamefont
  {Lange}},\ }\bibfield  {title} {\enquote {\bibinfo {title} {Floating surface
  potential of spherical dust grains in magnetized plasmas},}\ }\href@noop {}
  {\bibfield  {journal} {\bibinfo  {journal} {Journal of Plasma Physics}\
  }\textbf {\bibinfo {volume} {82}},\ \bibinfo {pages} {905820101} (\bibinfo
  {year} {2016})}\BibitemShut {NoStop}%
\bibitem [{\citenamefont {Patacchini}, \citenamefont {Hutchinson},\ and\
  \citenamefont {Lapenta}(2007)}]{dustcurrent}%
  \BibitemOpen
  \bibfield  {author} {\bibinfo {author} {\bibfnamefont {L.}~\bibnamefont
  {Patacchini}}, \bibinfo {author} {\bibfnamefont {I.~H.}\ \bibnamefont
  {Hutchinson}},\ and\ \bibinfo {author} {\bibfnamefont {G.}~\bibnamefont
  {Lapenta}},\ }\bibfield  {title} {\enquote {\bibinfo {title} {Electron
  collection by a negatively charged sphere in a collisionless
  magnetoplasma},}\ }\href@noop {} {\bibfield  {journal} {\bibinfo  {journal}
  {Physics of Plasmas}\ }\textbf {\bibinfo {volume} {14}},\ \bibinfo {pages}
  {062111} (\bibinfo {year} {2007})}\BibitemShut {NoStop}%
\bibitem [{\citenamefont {Yukihiro}\ \emph {et~al.}(2009)\citenamefont
  {Yukihiro}, \citenamefont {Gakushi}, \citenamefont {Takatoshi},\ and\
  \citenamefont {Osamu}}]{tomitadustchargingwithmagneticfield}%
  \BibitemOpen
  \bibfield  {author} {\bibinfo {author} {\bibfnamefont {T.}~\bibnamefont
  {Yukihiro}}, \bibinfo {author} {\bibfnamefont {K.}~\bibnamefont {Gakushi}},
  \bibinfo {author} {\bibfnamefont {Y.}~\bibnamefont {Takatoshi}},\ and\
  \bibinfo {author} {\bibfnamefont {I.}~\bibnamefont {Osamu}},\ }\bibfield
  {title} {\enquote {\bibinfo {title} {Charging of dust particles in magnetic
  field},}\ }\href@noop {} {\bibfield  {journal} {\bibinfo  {journal} {J.
  Plasma Fusion Res. SERIES}\ }\textbf {\bibinfo {volume} {8}},\ \bibinfo
  {pages} {273--276} (\bibinfo {year} {2009})}\BibitemShut {NoStop}%
\bibitem [{\citenamefont {{Kodanova}}\ \emph {et~al.}(2019)\citenamefont
  {{Kodanova}}, \citenamefont {{Bastykova}}, \citenamefont {{Ramazanov}},
  \citenamefont {{Nigmetova}}, \citenamefont {{Maiorov}},\ and\ \citenamefont
  {{Moldabekov}}}]{kodanovadustchargeb}%
  \BibitemOpen
  \bibfield  {author} {\bibinfo {author} {\bibfnamefont {S.~K.}\ \bibnamefont
  {{Kodanova}}}, \bibinfo {author} {\bibfnamefont {N.~K.}\ \bibnamefont
  {{Bastykova}}}, \bibinfo {author} {\bibfnamefont {T.~S.}\ \bibnamefont
  {{Ramazanov}}}, \bibinfo {author} {\bibfnamefont {G.~N.}\ \bibnamefont
  {{Nigmetova}}}, \bibinfo {author} {\bibfnamefont {S.~A.}\ \bibnamefont
  {{Maiorov}}},\ and\ \bibinfo {author} {\bibfnamefont {Z.~A.}\ \bibnamefont
  {{Moldabekov}}},\ }\bibfield  {title} {\enquote {\bibinfo {title} {Charging
  of a dust particle in a magnetized gas discharge plasma},}\ }\href@noop {}
  {\bibfield  {journal} {\bibinfo  {journal} {IEEE Transactions on Plasma
  Science}\ }\textbf {\bibinfo {volume} {47}},\ \bibinfo {pages} {3052--3056}
  (\bibinfo {year} {2019})}\BibitemShut {NoStop}%
\bibitem [{\citenamefont {Davari}, \citenamefont {Farokhi},\ and\ \citenamefont
  {Ali~Asgarian}(2023)}]{picsimulationmagnetizedcharge}%
  \BibitemOpen
  \bibfield  {author} {\bibinfo {author} {\bibfnamefont {H.}~\bibnamefont
  {Davari}}, \bibinfo {author} {\bibfnamefont {B.}~\bibnamefont {Farokhi}},\
  and\ \bibinfo {author} {\bibfnamefont {M.}~\bibnamefont {Ali~Asgarian}},\
  }\bibfield  {title} {\enquote {\bibinfo {title} {Particle simulation of the
  strong magnetic field effect on dust particle charging process},}\
  }\href@noop {} {\bibfield  {journal} {\bibinfo  {journal} {Scientific
  Reports}\ }\textbf {\bibinfo {volume} {13}},\ \bibinfo {pages} {197--203}
  (\bibinfo {year} {2023})}\BibitemShut {NoStop}%
\bibitem [{\citenamefont {Barkan}, \citenamefont {Merlino},\ and\ \citenamefont
  {D'Angelo}(1995)}]{daw2}%
  \BibitemOpen
  \bibfield  {author} {\bibinfo {author} {\bibfnamefont {A.}~\bibnamefont
  {Barkan}}, \bibinfo {author} {\bibfnamefont {R.~L.}\ \bibnamefont
  {Merlino}},\ and\ \bibinfo {author} {\bibfnamefont {N.}~\bibnamefont
  {D'Angelo}},\ }\bibfield  {title} {\enquote {\bibinfo {title} {Laboratory
  observation of the dust-acoustic wave mode},}\ }\href@noop {} {\bibfield
  {journal} {\bibinfo  {journal} {Phys. Plasmas}\ }\textbf {\bibinfo {volume}
  {2}},\ \bibinfo {pages} {3563--3565} (\bibinfo {year} {1995})}\BibitemShut
  {NoStop}%
\bibitem [{\citenamefont {Thomas}\ \emph
  {et~al.}(1994{\natexlab{b}})\citenamefont {Thomas}, \citenamefont {Morfill},
  \citenamefont {Demmel}, \citenamefont {Goree}, \citenamefont {Feuerbacher},\
  and\ \citenamefont {M\"ohlmann}}]{rfdischarge}%
  \BibitemOpen
  \bibfield  {author} {\bibinfo {author} {\bibfnamefont {H.}~\bibnamefont
  {Thomas}}, \bibinfo {author} {\bibfnamefont {G.~E.}\ \bibnamefont {Morfill}},
  \bibinfo {author} {\bibfnamefont {V.}~\bibnamefont {Demmel}}, \bibinfo
  {author} {\bibfnamefont {J.}~\bibnamefont {Goree}}, \bibinfo {author}
  {\bibfnamefont {B.}~\bibnamefont {Feuerbacher}},\ and\ \bibinfo {author}
  {\bibfnamefont {D.}~\bibnamefont {M\"ohlmann}},\ }\bibfield  {title}
  {\enquote {\bibinfo {title} {Plasma crystal: Coulomb crystallization in a
  dusty plasma},}\ }\href@noop {} {\bibfield  {journal} {\bibinfo  {journal}
  {Phys. Rev. Lett.}\ }\textbf {\bibinfo {volume} {73}},\ \bibinfo {pages}
  {652--655} (\bibinfo {year} {1994}{\natexlab{b}})}\BibitemShut {NoStop}%
\bibitem [{\citenamefont {Fortov}\ \emph {et~al.}(2003)\citenamefont {Fortov},
  \citenamefont {Usachev}, \citenamefont {Zobnin}, \citenamefont {Molotkov},\
  and\ \citenamefont {Petrov}}]{icpddw}%
  \BibitemOpen
  \bibfield  {author} {\bibinfo {author} {\bibfnamefont {V.~E.}\ \bibnamefont
  {Fortov}}, \bibinfo {author} {\bibfnamefont {A.~D.}\ \bibnamefont {Usachev}},
  \bibinfo {author} {\bibfnamefont {A.~V.}\ \bibnamefont {Zobnin}}, \bibinfo
  {author} {\bibfnamefont {V.~I.}\ \bibnamefont {Molotkov}},\ and\ \bibinfo
  {author} {\bibfnamefont {O.~F.}\ \bibnamefont {Petrov}},\ }\bibfield  {title}
  {\enquote {\bibinfo {title} {Dust-acoustic wave instability at the diffuse
  edge of radio frequency inductive low-pressure gas discharge plasma},}\
  }\href@noop {} {\bibfield  {journal} {\bibinfo  {journal} {Phys. Plasmas}\
  }\textbf {\bibinfo {volume} {10}},\ \bibinfo {pages} {1199--1208} (\bibinfo
  {year} {2003})}\BibitemShut {NoStop}%
\bibitem [{\citenamefont {Pramanik}\ \emph {et~al.}(2002)\citenamefont
  {Pramanik}, \citenamefont {Prasad}, \citenamefont {Sen},\ and\ \citenamefont
  {Kaw}}]{tsw}%
  \BibitemOpen
  \bibfield  {author} {\bibinfo {author} {\bibfnamefont {J.}~\bibnamefont
  {Pramanik}}, \bibinfo {author} {\bibfnamefont {G.}~\bibnamefont {Prasad}},
  \bibinfo {author} {\bibfnamefont {A.}~\bibnamefont {Sen}},\ and\ \bibinfo
  {author} {\bibfnamefont {P.~K.}\ \bibnamefont {Kaw}},\ }\bibfield  {title}
  {\enquote {\bibinfo {title} {Experimental observations of transverse shear
  waves in strongly coupled dusty plasmas},}\ }\href@noop {} {\bibfield
  {journal} {\bibinfo  {journal} {Phys. Rev. Lett.}\ }\textbf {\bibinfo
  {volume} {88}},\ \bibinfo {pages} {175001} (\bibinfo {year}
  {2002})}\BibitemShut {NoStop}%
\bibitem [{\citenamefont {Homann}\ \emph {et~al.}(1998)\citenamefont {Homann},
  \citenamefont {Melzer}, \citenamefont {Peters}, \citenamefont {Madani},\ and\
  \citenamefont {Piel}}]{dlw2}%
  \BibitemOpen
  \bibfield  {author} {\bibinfo {author} {\bibfnamefont {A.}~\bibnamefont
  {Homann}}, \bibinfo {author} {\bibfnamefont {A.}~\bibnamefont {Melzer}},
  \bibinfo {author} {\bibfnamefont {S.}~\bibnamefont {Peters}}, \bibinfo
  {author} {\bibfnamefont {R.}~\bibnamefont {Madani}},\ and\ \bibinfo {author}
  {\bibfnamefont {A.}~\bibnamefont {Piel}},\ }\bibfield  {title} {\enquote
  {\bibinfo {title} {Laser-excited dust lattice waves in plasma crystals},}\
  }\href@noop {} {\bibfield  {journal} {\bibinfo  {journal} {Physics Letters
  A}\ }\textbf {\bibinfo {volume} {242}},\ \bibinfo {pages} {173 -- 180}
  (\bibinfo {year} {1998})}\BibitemShut {NoStop}%
\bibitem [{\citenamefont {Rao}, \citenamefont {Shukla},\ and\ \citenamefont
  {Yu}(1990)}]{raodaw1}%
  \BibitemOpen
  \bibfield  {author} {\bibinfo {author} {\bibfnamefont {N.~N.}\ \bibnamefont
  {Rao}}, \bibinfo {author} {\bibfnamefont {P.~K.}\ \bibnamefont {Shukla}},\
  and\ \bibinfo {author} {\bibfnamefont {M.~Y.}\ \bibnamefont {Yu}},\
  }\bibfield  {title} {\enquote {\bibinfo {title} {Dust-acoustic waves in dusty
  plasmas},}\ }\href@noop {} {\bibfield  {journal} {\bibinfo  {journal}
  {Planet. Space Sci.}\ }\textbf {\bibinfo {volume} {38}},\ \bibinfo {pages}
  {543--546} (\bibinfo {year} {1990})}\BibitemShut {NoStop}%
\bibitem [{\citenamefont {Farokhi}\ \emph {et~al.}(2000)\citenamefont
  {Farokhi}, \citenamefont {Shukla}, \citenamefont {Tsintsadze},\ and\
  \citenamefont {Tskhakaya}}]{dlw1}%
  \BibitemOpen
  \bibfield  {author} {\bibinfo {author} {\bibfnamefont {B.}~\bibnamefont
  {Farokhi}}, \bibinfo {author} {\bibfnamefont {P.~K.}\ \bibnamefont {Shukla}},
  \bibinfo {author} {\bibfnamefont {N.~L.}\ \bibnamefont {Tsintsadze}},\ and\
  \bibinfo {author} {\bibfnamefont {D.~D.}\ \bibnamefont {Tskhakaya}},\
  }\bibfield  {title} {\enquote {\bibinfo {title} {Dust lattice waves in a
  plasma crystal},}\ }\href@noop {} {\bibfield  {journal} {\bibinfo  {journal}
  {Phys. Plasmas}\ }\textbf {\bibinfo {volume} {7}},\ \bibinfo {pages}
  {814--818} (\bibinfo {year} {2000})}\BibitemShut {NoStop}%
\bibitem [{\citenamefont {Choudhary}\ \emph
  {et~al.}(2020{\natexlab{b}})\citenamefont {Choudhary}, \citenamefont
  {Bergert}, \citenamefont {Mitic},\ and\ \citenamefont
  {Thoma}}]{mangilalmagneticdaw}%
  \BibitemOpen
  \bibfield  {author} {\bibinfo {author} {\bibfnamefont {M.}~\bibnamefont
  {Choudhary}}, \bibinfo {author} {\bibfnamefont {R.}~\bibnamefont {Bergert}},
  \bibinfo {author} {\bibfnamefont {S.}~\bibnamefont {Mitic}},\ and\ \bibinfo
  {author} {\bibfnamefont {M.~H.}\ \bibnamefont {Thoma}},\ }\bibfield  {title}
  {\enquote {\bibinfo {title} {Influence of external magnetic field on dust
  acoustic waves in a capacitive rf discharge},}\ }\href@noop {} {\bibfield
  {journal} {\bibinfo  {journal} {Contributions to Plasma Physics}\ }\textbf
  {\bibinfo {volume} {60}},\ \bibinfo {pages} {e201900115} (\bibinfo {year}
  {2020}{\natexlab{b}})}\BibitemShut {NoStop}%
\bibitem [{\citenamefont {Melzer}\ \emph {et~al.}(2020)\citenamefont {Melzer},
  \citenamefont {Krüger}, \citenamefont {Schütt},\ and\ \citenamefont
  {Maier}}]{melzerdaw}%
  \BibitemOpen
  \bibfield  {author} {\bibinfo {author} {\bibfnamefont {A.}~\bibnamefont
  {Melzer}}, \bibinfo {author} {\bibfnamefont {H.}~\bibnamefont {Krüger}},
  \bibinfo {author} {\bibfnamefont {S.}~\bibnamefont {Schütt}},\ and\ \bibinfo
  {author} {\bibfnamefont {D.}~\bibnamefont {Maier}},\ }\bibfield  {title}
  {\enquote {\bibinfo {title} {Dust-density waves in radio-frequency discharges
  under magnetic fields},}\ }\href@noop {} {\bibfield  {journal} {\bibinfo
  {journal} {Physics of Plasmas}\ }\textbf {\bibinfo {volume} {27}},\ \bibinfo
  {pages} {033704} (\bibinfo {year} {2020})}\BibitemShut {NoStop}%
\bibitem [{\citenamefont {Salimullah}\ and\ \citenamefont
  {Salahuddin}(1998)}]{dawmagnetized1}%
  \BibitemOpen
  \bibfield  {author} {\bibinfo {author} {\bibfnamefont {M.}~\bibnamefont
  {Salimullah}}\ and\ \bibinfo {author} {\bibfnamefont {M.}~\bibnamefont
  {Salahuddin}},\ }\bibfield  {title} {\enquote {\bibinfo {title}
  {Dust-acoustic waves in a magnetized dusty plasma},}\ }\href@noop {}
  {\bibfield  {journal} {\bibinfo  {journal} {Physics of Plasmas}\ }\textbf
  {\bibinfo {volume} {5}},\ \bibinfo {pages} {828--829} (\bibinfo {year}
  {1998})}\BibitemShut {NoStop}%
\bibitem [{\citenamefont {Agarwal}\ and\ \citenamefont
  {Prasad}(2003)}]{agarwalrotation}%
  \BibitemOpen
  \bibfield  {author} {\bibinfo {author} {\bibfnamefont {A.}~\bibnamefont
  {Agarwal}}\ and\ \bibinfo {author} {\bibfnamefont {G.}~\bibnamefont
  {Prasad}},\ }\bibfield  {title} {\enquote {\bibinfo {title} {Spontaneous dust
  mass rotation in an unmagnetized dusty plasma},}\ }\href@noop {} {\bibfield
  {journal} {\bibinfo  {journal} {Physics Letters A}\ }\textbf {\bibinfo
  {volume} {309}},\ \bibinfo {pages} {103 --108} (\bibinfo {year}
  {2003})}\BibitemShut {NoStop}%
\bibitem [{\citenamefont {Kaur}\ \emph {et~al.}(2015)\citenamefont {Kaur},
  \citenamefont {Bose}, \citenamefont {Chattopadhyay}, \citenamefont {Sharma},
  \citenamefont {Ghosh}, \citenamefont {Saxena},\ and\ \citenamefont
  {Thomas}}]{manjeetrotation}%
  \BibitemOpen
  \bibfield  {author} {\bibinfo {author} {\bibfnamefont {M.}~\bibnamefont
  {Kaur}}, \bibinfo {author} {\bibfnamefont {S.}~\bibnamefont {Bose}}, \bibinfo
  {author} {\bibfnamefont {P.~K.}\ \bibnamefont {Chattopadhyay}}, \bibinfo
  {author} {\bibfnamefont {D.}~\bibnamefont {Sharma}}, \bibinfo {author}
  {\bibfnamefont {J.}~\bibnamefont {Ghosh}}, \bibinfo {author} {\bibfnamefont
  {Y.~C.}\ \bibnamefont {Saxena}},\ and\ \bibinfo {author} {\bibfnamefont
  {E.}~\bibnamefont {Thomas}},\ }\bibfield  {title} {\enquote {\bibinfo {title}
  {Generation of multiple toroidal dust vortices by a non-monotonic density
  gradient in a direct current glow discharge plasma},}\ }\href@noop {}
  {\bibfield  {journal} {\bibinfo  {journal} {Phys. Plasmas}\ }\textbf
  {\bibinfo {volume} {22}},\ \bibinfo {pages} {093702} (\bibinfo {year}
  {2015})}\BibitemShut {NoStop}%
\bibitem [{\citenamefont {Mitic}\ \emph {et~al.}(2008)\citenamefont {Mitic},
  \citenamefont {S\"utterlin}, \citenamefont {H\"ofner}, \citenamefont {Thoma},
  \citenamefont {Zhdanov},\ and\ \citenamefont
  {Morfill}}]{thermalcreeprotation}%
  \BibitemOpen
  \bibfield  {author} {\bibinfo {author} {\bibfnamefont {S.}~\bibnamefont
  {Mitic}}, \bibinfo {author} {\bibfnamefont {R.}~\bibnamefont {S\"utterlin}},
  \bibinfo {author} {\bibfnamefont {A.~V. I.~H.}\ \bibnamefont {H\"ofner}},
  \bibinfo {author} {\bibfnamefont {M.~H.}\ \bibnamefont {Thoma}}, \bibinfo
  {author} {\bibfnamefont {S.}~\bibnamefont {Zhdanov}},\ and\ \bibinfo {author}
  {\bibfnamefont {G.~E.}\ \bibnamefont {Morfill}},\ }\bibfield  {title}
  {\enquote {\bibinfo {title} {Convective dust clouds driven by thermal creep
  in a complex plasma},}\ }\href@noop {} {\bibfield  {journal} {\bibinfo
  {journal} {Phys. Rev. Lett.}\ }\textbf {\bibinfo {volume} {101}},\ \bibinfo
  {pages} {235001} (\bibinfo {year} {2008})}\BibitemShut {NoStop}%
\bibitem [{\citenamefont {Konopka}\ \emph {et~al.}(2000)\citenamefont
  {Konopka}, \citenamefont {Samsonov}, \citenamefont {Ivlev}, \citenamefont
  {Goree}, \citenamefont {Steinberg},\ and\ \citenamefont
  {Morfill}}]{rotationknopka1}%
  \BibitemOpen
  \bibfield  {author} {\bibinfo {author} {\bibfnamefont {U.}~\bibnamefont
  {Konopka}}, \bibinfo {author} {\bibfnamefont {D.}~\bibnamefont {Samsonov}},
  \bibinfo {author} {\bibfnamefont {A.~V.}\ \bibnamefont {Ivlev}}, \bibinfo
  {author} {\bibfnamefont {J.}~\bibnamefont {Goree}}, \bibinfo {author}
  {\bibfnamefont {V.}~\bibnamefont {Steinberg}},\ and\ \bibinfo {author}
  {\bibfnamefont {G.~E.}\ \bibnamefont {Morfill}},\ }\bibfield  {title}
  {\enquote {\bibinfo {title} {Rigid and differential plasma crystal rotation
  induced by magnetic fields},}\ }\href@noop {} {\bibfield  {journal} {\bibinfo
   {journal} {Phys. Rev. E}\ }\textbf {\bibinfo {volume} {61}},\ \bibinfo
  {pages} {1890--1898} (\bibinfo {year} {2000})}\BibitemShut {NoStop}%
\bibitem [{\citenamefont {Feng}\ \emph {et~al.}(2013)\citenamefont {Feng},
  \citenamefont {Yan-Hong}, \citenamefont {Zhao-Yang}, \citenamefont {Long},\
  and\ \citenamefont {Mao-Fu}}]{clusterrotationunmagnetizedplasma}%
  \BibitemOpen
  \bibfield  {author} {\bibinfo {author} {\bibfnamefont {H.}~\bibnamefont
  {Feng}}, \bibinfo {author} {\bibfnamefont {L.}~\bibnamefont {Yan-Hong}},
  \bibinfo {author} {\bibfnamefont {C.}~\bibnamefont {Zhao-Yang}}, \bibinfo
  {author} {\bibfnamefont {W.}~\bibnamefont {Long}},\ and\ \bibinfo {author}
  {\bibfnamefont {Y.}~\bibnamefont {Mao-Fu}},\ }\bibfield  {title} {\enquote
  {\bibinfo {title} {Cluster rotation in an unmagnetized dusty plasma},}\
  }\href@noop {} {\bibfield  {journal} {\bibinfo  {journal} {Chinese Physics
  Letters}\ }\textbf {\bibinfo {volume} {30}},\ \bibinfo {pages} {115201}
  (\bibinfo {year} {2013})}\BibitemShut {NoStop}%
\bibitem [{\citenamefont {Cheung}\ \emph {et~al.}(2003)\citenamefont {Cheung},
  \citenamefont {Prior}, \citenamefont {Mitchell}, \citenamefont {Samarian},\
  and\ \citenamefont {James}}]{inductivelycoupledrotation}%
  \BibitemOpen
  \bibfield  {author} {\bibinfo {author} {\bibfnamefont {F.~M.~H.}\
  \bibnamefont {Cheung}}, \bibinfo {author} {\bibfnamefont {N.~J.}\
  \bibnamefont {Prior}}, \bibinfo {author} {\bibfnamefont {L.~W.}\ \bibnamefont
  {Mitchell}}, \bibinfo {author} {\bibfnamefont {A.~A.}\ \bibnamefont
  {Samarian}},\ and\ \bibinfo {author} {\bibfnamefont {B.~W.}\ \bibnamefont
  {James}},\ }\bibfield  {title} {\enquote {\bibinfo {title} {Rotation of
  coulomb crystals in a magnetized inductively coupled complex plasma},}\
  }\href@noop {} {\bibfield  {journal} {\bibinfo  {journal} {IEEE Transactions
  on Plasma Science}\ }\textbf {\bibinfo {volume} {31}},\ \bibinfo {pages}
  {112--118} (\bibinfo {year} {2003})}\BibitemShut {NoStop}%
\bibitem [{\citenamefont {Ishihara}\ and\ \citenamefont
  {Sato}(2001)}]{rotationinionflow}%
  \BibitemOpen
  \bibfield  {author} {\bibinfo {author} {\bibfnamefont {O.}~\bibnamefont
  {Ishihara}}\ and\ \bibinfo {author} {\bibfnamefont {N.}~\bibnamefont
  {Sato}},\ }\bibfield  {title} {\enquote {\bibinfo {title} {On the rotation of
  a dust particulate in an ion flow in a magnetic field},}\ }\href@noop {}
  {\bibfield  {journal} {\bibinfo  {journal} {IEEE Transactions on Plasma
  Science}\ }\textbf {\bibinfo {volume} {29}},\ \bibinfo {pages} {179--181}
  (\bibinfo {year} {2001})}\BibitemShut {NoStop}%
\bibitem [{\citenamefont {Choudhary}\ \emph
  {et~al.}(2020{\natexlab{c}})\citenamefont {Choudhary}, \citenamefont
  {Bergert}, \citenamefont {Mitic},\ and\ \citenamefont
  {Thoma}}]{mangimagneticrotation}%
  \BibitemOpen
  \bibfield  {author} {\bibinfo {author} {\bibfnamefont {M.}~\bibnamefont
  {Choudhary}}, \bibinfo {author} {\bibfnamefont {R.}~\bibnamefont {Bergert}},
  \bibinfo {author} {\bibfnamefont {S.}~\bibnamefont {Mitic}},\ and\ \bibinfo
  {author} {\bibfnamefont {M.~H.}\ \bibnamefont {Thoma}},\ }\bibfield  {title}
  {\enquote {\bibinfo {title} {Three-dimensional dusty plasma in a strong
  magnetic field: Observation of rotating dust tori},}\ }\href@noop {}
  {\bibfield  {journal} {\bibinfo  {journal} {Physics of Plasmas}\ }\textbf
  {\bibinfo {volume} {27}},\ \bibinfo {pages} {063701} (\bibinfo {year}
  {2020}{\natexlab{c}})}\BibitemShut {NoStop}%
\bibitem [{\citenamefont {Law}\ \emph {et~al.}(1998)\citenamefont {Law},
  \citenamefont {Steel}, \citenamefont {Annaratone},\ and\ \citenamefont
  {Allen}}]{circulation}%
  \BibitemOpen
  \bibfield  {author} {\bibinfo {author} {\bibfnamefont {D.~A.}\ \bibnamefont
  {Law}}, \bibinfo {author} {\bibfnamefont {W.~H.}\ \bibnamefont {Steel}},
  \bibinfo {author} {\bibfnamefont {B.~M.}\ \bibnamefont {Annaratone}},\ and\
  \bibinfo {author} {\bibfnamefont {J.~E.}\ \bibnamefont {Allen}},\ }\bibfield
  {title} {\enquote {\bibinfo {title} {Probe-induced particle circulation in a
  plasma crystal},}\ }\href@noop {} {\bibfield  {journal} {\bibinfo  {journal}
  {Phys. Rev. Lett.}\ }\textbf {\bibinfo {volume} {80}},\ \bibinfo {pages}
  {4189--4192} (\bibinfo {year} {1998})}\BibitemShut {NoStop}%
\bibitem [{\citenamefont {Kaw}, \citenamefont {Nishikawa},\ and\ \citenamefont
  {Sato}(2002)}]{kawmodelrotation}%
  \BibitemOpen
  \bibfield  {author} {\bibinfo {author} {\bibfnamefont {P.~K.}\ \bibnamefont
  {Kaw}}, \bibinfo {author} {\bibfnamefont {K.}~\bibnamefont {Nishikawa}},\
  and\ \bibinfo {author} {\bibfnamefont {N.}~\bibnamefont {Sato}},\ }\bibfield
  {title} {\enquote {\bibinfo {title} {Rotation in collisional strongly coupled
  dusty plasmas in a magnetic field},}\ }\href@noop {} {\bibfield  {journal}
  {\bibinfo  {journal} {Physics of Plasmas}\ }\textbf {\bibinfo {volume} {9}},\
  \bibinfo {pages} {387--390} (\bibinfo {year} {2002})}\BibitemShut {NoStop}%
\bibitem [{\citenamefont {Vaulina}\ \emph {et~al.}(2001)\citenamefont
  {Vaulina}, \citenamefont {Samarian}, \citenamefont {Nefedov},\ and\
  \citenamefont {Fortov}}]{selfexcitedmotioninhomogeneus}%
  \BibitemOpen
  \bibfield  {author} {\bibinfo {author} {\bibfnamefont {O.}~\bibnamefont
  {Vaulina}}, \bibinfo {author} {\bibfnamefont {A.}~\bibnamefont {Samarian}},
  \bibinfo {author} {\bibfnamefont {A.}~\bibnamefont {Nefedov}},\ and\ \bibinfo
  {author} {\bibfnamefont {V.}~\bibnamefont {Fortov}},\ }\bibfield  {title}
  {\enquote {\bibinfo {title} {Self-excited motion of dust particles in a
  inhomogeneous plasma},}\ }\href@noop {} {\bibfield  {journal} {\bibinfo
  {journal} {Physics Letters A}\ }\textbf {\bibinfo {volume} {289}},\ \bibinfo
  {pages} {240--244} (\bibinfo {year} {2001})}\BibitemShut {NoStop}%
\bibitem [{\citenamefont {Laishram}, \citenamefont {Sharma},\ and\
  \citenamefont {Kaw}(2014)}]{laishramshearflowexplaination}%
  \BibitemOpen
  \bibfield  {author} {\bibinfo {author} {\bibfnamefont {M.}~\bibnamefont
  {Laishram}}, \bibinfo {author} {\bibfnamefont {D.}~\bibnamefont {Sharma}},\
  and\ \bibinfo {author} {\bibfnamefont {P.~K.}\ \bibnamefont {Kaw}},\
  }\bibfield  {title} {\enquote {\bibinfo {title} {Dynamics of a confined dusty
  fluid in a sheared ion flow},}\ }\href@noop {} {\bibfield  {journal}
  {\bibinfo  {journal} {Phys. Plasmas}\ }\textbf {\bibinfo {volume} {21}},\
  \bibinfo {pages} {073703} (\bibinfo {year} {2014})}\BibitemShut {NoStop}%
\bibitem [{\citenamefont {Akdim}\ and\ \citenamefont
  {Goedheer}(2003)}]{modelingdustvortices}%
  \BibitemOpen
  \bibfield  {author} {\bibinfo {author} {\bibfnamefont {M.~R.}\ \bibnamefont
  {Akdim}}\ and\ \bibinfo {author} {\bibfnamefont {W.~J.}\ \bibnamefont
  {Goedheer}},\ }\bibfield  {title} {\enquote {\bibinfo {title} {Modeling of
  self-excited dust vortices in complex plasmas under microgravity},}\
  }\href@noop {} {\bibfield  {journal} {\bibinfo  {journal} {Phys. Rev. E}\
  }\textbf {\bibinfo {volume} {67}},\ \bibinfo {pages} {056405} (\bibinfo
  {year} {2003})}\BibitemShut {NoStop}%
\bibitem [{\citenamefont {Klindworth}\ \emph {et~al.}(2004)\citenamefont
  {Klindworth}, \citenamefont {Piel}, \citenamefont {Melzer}, \citenamefont
  {Konopka}, \citenamefont {Rothermel}, \citenamefont {Tarantik},\ and\
  \citenamefont {Morfill}}]{microgravityvoid}%
  \BibitemOpen
  \bibfield  {author} {\bibinfo {author} {\bibfnamefont {M.}~\bibnamefont
  {Klindworth}}, \bibinfo {author} {\bibfnamefont {A.}~\bibnamefont {Piel}},
  \bibinfo {author} {\bibfnamefont {A.}~\bibnamefont {Melzer}}, \bibinfo
  {author} {\bibfnamefont {U.}~\bibnamefont {Konopka}}, \bibinfo {author}
  {\bibfnamefont {H.}~\bibnamefont {Rothermel}}, \bibinfo {author}
  {\bibfnamefont {K.}~\bibnamefont {Tarantik}},\ and\ \bibinfo {author}
  {\bibfnamefont {G.~E.}\ \bibnamefont {Morfill}},\ }\bibfield  {title}
  {\enquote {\bibinfo {title} {Dust-free regions around langmuir probes in
  complex plasmas under microgravity},}\ }\href@noop {} {\bibfield  {journal}
  {\bibinfo  {journal} {Phys. Rev. Lett.}\ }\textbf {\bibinfo {volume} {93}},\
  \bibinfo {pages} {195002} (\bibinfo {year} {2004})}\BibitemShut {NoStop}%
\bibitem [{\citenamefont {Goree}\ \emph {et~al.}(1999)\citenamefont {Goree},
  \citenamefont {Morfill}, \citenamefont {Tsytovich},\ and\ \citenamefont
  {Vladimirov}}]{voidstheorygoree}%
  \BibitemOpen
  \bibfield  {author} {\bibinfo {author} {\bibfnamefont {J.}~\bibnamefont
  {Goree}}, \bibinfo {author} {\bibfnamefont {G.~E.}\ \bibnamefont {Morfill}},
  \bibinfo {author} {\bibfnamefont {V.~N.}\ \bibnamefont {Tsytovich}},\ and\
  \bibinfo {author} {\bibfnamefont {S.~V.}\ \bibnamefont {Vladimirov}},\
  }\bibfield  {title} {\enquote {\bibinfo {title} {Theory of dust voids in
  plasmas},}\ }\href@noop {} {\bibfield  {journal} {\bibinfo  {journal} {Phys
  Rev E Stat Phys Plasmas Fluids Relat Interdiscip Topics.}\ }\textbf {\bibinfo
  {volume} {59}},\ \bibinfo {pages} {7055--67} (\bibinfo {year}
  {1999})}\BibitemShut {NoStop}%
\bibitem [{\citenamefont {Vasilyak}, \citenamefont {Vetchinin},\ and\
  \citenamefont {Polyakov}(2023)}]{voiddependsonionization}%
  \BibitemOpen
  \bibfield  {author} {\bibinfo {author} {\bibfnamefont {L.~M.}\ \bibnamefont
  {Vasilyak}}, \bibinfo {author} {\bibfnamefont {S.~P.}\ \bibnamefont
  {Vetchinin}},\ and\ \bibinfo {author} {\bibfnamefont {D.~N.}\ \bibnamefont
  {Polyakov}},\ }\bibfield  {title} {\enquote {\bibinfo {title} {Effect of
  ionization on void formation in an rf discharge under microgravity
  conditions},}\ }\href@noop {} {\bibfield  {journal} {\bibinfo  {journal}
  {Plasma Physics Reports}\ }\textbf {\bibinfo {volume} {49}},\ \bibinfo
  {pages} {290–295} (\bibinfo {year} {2023})}\BibitemShut {NoStop}%
\bibitem [{\citenamefont {Donkó}, \citenamefont {Hartmann},\ and\
  \citenamefont {Kalman}(2009)}]{crystaldonko2009}%
  \BibitemOpen
  \bibfield  {author} {\bibinfo {author} {\bibfnamefont {Z.}~\bibnamefont
  {Donkó}}, \bibinfo {author} {\bibfnamefont {P.}~\bibnamefont {Hartmann}},\
  and\ \bibinfo {author} {\bibfnamefont {G.~J.}\ \bibnamefont {Kalman}},\
  }\bibfield  {title} {\enquote {\bibinfo {title} {Two-dimensional dusty plasma
  crystals and liquids},}\ }\href@noop {} {\bibfield  {journal} {\bibinfo
  {journal} {Journal of Physics: Conference Series}\ }\textbf {\bibinfo
  {volume} {162}},\ \bibinfo {pages} {012016} (\bibinfo {year}
  {2009})}\BibitemShut {NoStop}%
\bibitem [{\citenamefont {Arumugam}\ \emph {et~al.}(2021)\citenamefont
  {Arumugam}, \citenamefont {Bandyopadhyay}, \citenamefont {Singh},
  \citenamefont {Hariprasad}, \citenamefont {Rathod}, \citenamefont {Arora},\
  and\ \citenamefont {Sen}}]{crystalarumugamdc2021}%
  \BibitemOpen
  \bibfield  {author} {\bibinfo {author} {\bibfnamefont {S.}~\bibnamefont
  {Arumugam}}, \bibinfo {author} {\bibfnamefont {P.}~\bibnamefont
  {Bandyopadhyay}}, \bibinfo {author} {\bibfnamefont {S.}~\bibnamefont
  {Singh}}, \bibinfo {author} {\bibfnamefont {M.~G.}\ \bibnamefont
  {Hariprasad}}, \bibinfo {author} {\bibfnamefont {D.}~\bibnamefont {Rathod}},
  \bibinfo {author} {\bibfnamefont {G.}~\bibnamefont {Arora}},\ and\ \bibinfo
  {author} {\bibfnamefont {A.}~\bibnamefont {Sen}},\ }\bibfield  {title}
  {\enquote {\bibinfo {title} {Dpex-ii: a new dusty plasma device capable of
  producing large sized dc coulomb crystals},}\ }\href@noop {} {\bibfield
  {journal} {\bibinfo  {journal} {Plasma Sources Science and Technology}\
  }\textbf {\bibinfo {volume} {30}},\ \bibinfo {pages} {085003} (\bibinfo
  {year} {2021})}\BibitemShut {NoStop}%
\bibitem [{\citenamefont {Chaubey}\ and\ \citenamefont
  {Goree}(2022{\natexlab{b}})}]{crystalneeraj2023}%
  \BibitemOpen
  \bibfield  {author} {\bibinfo {author} {\bibfnamefont {N.}~\bibnamefont
  {Chaubey}}\ and\ \bibinfo {author} {\bibfnamefont {J.}~\bibnamefont
  {Goree}},\ }\bibfield  {title} {\enquote {\bibinfo {title} {Preservation of a
  dust crystal as it falls in an afterglow plasma},}\ }\href@noop {} {\bibfield
   {journal} {\bibinfo  {journal} {Frontiers in Physics}\ }\textbf {\bibinfo
  {volume} {10}},\ \bibinfo {pages} {879092} (\bibinfo {year}
  {2022}{\natexlab{b}})}\BibitemShut {NoStop}%
\bibitem [{\citenamefont {Jaiswal}\ \emph {et~al.}(2017)\citenamefont
  {Jaiswal}, \citenamefont {Hall}, \citenamefont {LeBlanc}, \citenamefont
  {Mukherjee},\ and\ \citenamefont {Thomas}}]{crystalmagneticfieldsurbhi2017}%
  \BibitemOpen
  \bibfield  {author} {\bibinfo {author} {\bibfnamefont {S.}~\bibnamefont
  {Jaiswal}}, \bibinfo {author} {\bibfnamefont {T.}~\bibnamefont {Hall}},
  \bibinfo {author} {\bibfnamefont {S.}~\bibnamefont {LeBlanc}}, \bibinfo
  {author} {\bibfnamefont {R.}~\bibnamefont {Mukherjee}},\ and\ \bibinfo
  {author} {\bibfnamefont {E.}~\bibnamefont {Thomas}},\ }\bibfield  {title}
  {\enquote {\bibinfo {title} {Effect of magnetic field on the phase transition
  in a dusty plasma},}\ }\href@noop {} {\bibfield  {journal} {\bibinfo
  {journal} {Physics of Plasmas}\ }\textbf {\bibinfo {volume} {24}},\ \bibinfo
  {pages} {113703} (\bibinfo {year} {2017})}\BibitemShut {NoStop}%
\bibitem [{\citenamefont {Thomas}, \citenamefont {Merlino},\ and\ \citenamefont
  {Rosenberg}(2012)}]{thomasmagnetizationdust}%
  \BibitemOpen
  \bibfield  {author} {\bibinfo {author} {\bibfnamefont {E.}~\bibnamefont
  {Thomas}}, \bibinfo {author} {\bibfnamefont {R.~L.}\ \bibnamefont
  {Merlino}},\ and\ \bibinfo {author} {\bibfnamefont {M.}~\bibnamefont
  {Rosenberg}},\ }\bibfield  {title} {\enquote {\bibinfo {title} {Magnetized
  dusty plasmas: the next frontier for complex plasma research},}\ }\href@noop
  {} {\bibfield  {journal} {\bibinfo  {journal} {Plasma Physics and Controlled
  Fusion}\ }\textbf {\bibinfo {volume} {54}},\ \bibinfo {pages} {124034}
  (\bibinfo {year} {2012})}\BibitemShut {NoStop}%
\bibitem [{\citenamefont {Melzer}\ \emph {et~al.}(2021)\citenamefont {Melzer},
  \citenamefont {Krüger}, \citenamefont {Maier},\ and\ \citenamefont
  {Schütt}}]{melzermagnetizationpaper}%
  \BibitemOpen
  \bibfield  {author} {\bibinfo {author} {\bibfnamefont {A.}~\bibnamefont
  {Melzer}}, \bibinfo {author} {\bibfnamefont {H.}~\bibnamefont {Krüger}},
  \bibinfo {author} {\bibfnamefont {D.}~\bibnamefont {Maier}},\ and\ \bibinfo
  {author} {\bibfnamefont {S.}~\bibnamefont {Schütt}},\ }\bibfield  {title}
  {\enquote {\bibinfo {title} {Physics of magnetized dusty plasmas},}\
  }\href@noop {} {\bibfield  {journal} {\bibinfo  {journal} {Reviews of Modern
  Plasma Physics}\ }\textbf {\bibinfo {volume} {5}},\ \bibinfo {pages} {11}
  (\bibinfo {year} {2021})}\BibitemShut {NoStop}%
\bibitem [{\citenamefont {Menati}, \citenamefont {Thomas},\ and\ \citenamefont
  {Kushner}(2019)}]{filamentmenati2020}%
  \BibitemOpen
  \bibfield  {author} {\bibinfo {author} {\bibfnamefont {M.}~\bibnamefont
  {Menati}}, \bibinfo {author} {\bibfnamefont {E.}~\bibnamefont {Thomas}},\
  and\ \bibinfo {author} {\bibfnamefont {M.~J.}\ \bibnamefont {Kushner}},\
  }\bibfield  {title} {\enquote {\bibinfo {title} {{Filamentation of
  capacitively coupled plasmas in large magnetic fields}},}\ }\href@noop {}
  {\bibfield  {journal} {\bibinfo  {journal} {Physics of Plasmas}\ }\textbf
  {\bibinfo {volume} {26}},\ \bibinfo {pages} {063515} (\bibinfo {year}
  {2019})}\BibitemShut {NoStop}%
\bibitem [{\citenamefont {Chaubey}\ \emph {et~al.}(2021)\citenamefont
  {Chaubey}, \citenamefont {Goree}, \citenamefont {Lanham},\ and\ \citenamefont
  {Kushner}}]{neerajpositivechargepop}%
  \BibitemOpen
  \bibfield  {author} {\bibinfo {author} {\bibfnamefont {N.}~\bibnamefont
  {Chaubey}}, \bibinfo {author} {\bibfnamefont {J.}~\bibnamefont {Goree}},
  \bibinfo {author} {\bibfnamefont {S.~J.}\ \bibnamefont {Lanham}},\ and\
  \bibinfo {author} {\bibfnamefont {M.~J.}\ \bibnamefont {Kushner}},\
  }\bibfield  {title} {\enquote {\bibinfo {title} {Positive charging of grains
  in an afterglow plasma is enhanced by ions drifting in an electric field},}\
  }\href@noop {} {\bibfield  {journal} {\bibinfo  {journal} {Physics of
  Plasmas}\ }\textbf {\bibinfo {volume} {28}},\ \bibinfo {pages} {103702}
  (\bibinfo {year} {2021})}\BibitemShut {NoStop}%
\bibitem [{\citenamefont {Chaubey}\ and\ \citenamefont
  {Goree}(2022{\natexlab{c}})}]{neerajpositivecrystal}%
  \BibitemOpen
  \bibfield  {author} {\bibinfo {author} {\bibfnamefont {N.}~\bibnamefont
  {Chaubey}}\ and\ \bibinfo {author} {\bibfnamefont {J.}~\bibnamefont
  {Goree}},\ }\bibfield  {title} {\enquote {\bibinfo {title} {Coulomb expansion
  of a thin dust cloud observed experimentally under afterglow plasma
  conditions},}\ }\href@noop {} {\bibfield  {journal} {\bibinfo  {journal}
  {Physics of Plasmas}\ }\textbf {\bibinfo {volume} {29}},\ \bibinfo {pages}
  {113705} (\bibinfo {year} {2022}{\natexlab{c}})}\BibitemShut {NoStop}%
\end{thebibliography}%
\end{document}